\documentclass[utf8]{FrontiersinVancouver}
\setcitestyle{square}
\usepackage{url,hyperref,lineno,microtype,silence}
\usepackage[onehalfspacing]{setspace}
\def\keyFont{\fontsize{8}{11}\helveticabold}
\def\firstAuthorLast{Szydagis {et~al.}}
\def\Authors{M.~Szydagis\,$^{1,*}$, J.~Balajthy\,$^{2,3}$, G.A.~Block\,$^{1,4,5}$, J.P.~Brodsky\,$^{6}$, E.~Brown\,$^{4}$, J.E.~Cutter\,$^{2,7}$, S.J.~Farrell\,$^{8}$, J.~Huang\,$^{2,9}$, A.C.~Kamaha\,$^{1,10}$, E.S.~Kozlova\,$^{11}$, C.S.~Liebenthal\,$^{8}$, D.N.~McKinsey\,$^{12,13}$, K.~McMichael\,$^{4}$, R.~McMonigle\,$^{1}$, M.~Mooney\,$^{14}$, J.~Mueller\,$^{14}$, K.~Ni\,$^{9}$, G.R.C.~Rischbieter\,$^{1,15,17,*}$, K.~Trengove\,$^{1,10}$, M.~Tripathi\,$^{2}$, C.D.~Tunnell\,$^{8}$, V.~Velan\,$^{12}$, S.~Westerdale\,$^{16}$, M.D.~Wyman\,$^{1}$, Z.~Zhao\,$^{9}$, and M.~Zhong\,$^{9}$}

\def\Address{$^{1}$Department of Physics, University at Albany, State University of New York, Albany, NY, USA \\
$^{2}$Department of Physics, University of California Davis, Davis, CA, USA \\
$^{3}$Sandia National Laboratories, Livermore, CA, USA \\
$^{4}$Department of Physics, Applied Physics and Astronomy, Rensselaer Polytechnic Institute, Troy, NY, USA \\
$^{5}$Department of Physics and Astronomy, University of New Mexico, Albuquerque, NM, USA \\
$^{6}$Lawrence Livermore National Laboratory, Livermore, CA, USA \\
$^{7}$Deepgram, Mountain View, CA, USA \\
$^{8}$Department of Physics and Astronomy, Rice University, Houston, TX, USA \\
$^{9}$Department of Physics, University of California San Diego, La Jolla, CA, USA \\
$^{10}$Department of Physics and Astronomy, University of California Los Angeles, Los Angeles, CA, USA \\
$^{11}$ORCID: 0000-0002-1976-3425\\
$^{12}$Lawrence Berkeley National Laboratory, Berkeley, CA, USA \\
$^{13}$Department of Physics, University of California Berkeley, Berkeley, CA, USA \\
$^{14}$Department of Physics, Colorado State University, Fort Collins, CO, USA \\
$^{15}$Department of Physics, University of Michigan, Ann Arbor, MI, USA \\
$^{16}$Department of Physics \& Astronomy, University of California Riverside, Riverside, CA, USA \\
$^{17}$Physik-Institut, University of Z{\"u}rich, Z{\"u}rich, Switzerland}

\def\corrAuthor{Matthew Szydagis \\ Department of Physics, \\ University at Albany, State University of New York, \\ 1400 Washington Av., Albany, NY 12222, USA}

\def\corrEmail{mszydagis@albany.edu}

\begin{document}
\onecolumn
\firstpage{1}

\title[A Review of NEST Models for Liquid Xenon]{A Review of NEST Models for Liquid Xenon \& Exhaustive Comparison to Other Approaches}

\author[\firstAuthorLast ]{\Authors}
\address{}
\correspondance{}

\extraAuth{Gregory R.C.~Rischbieter \\ Department of Physics, University of Michigan, \\ 450 Church Street, Ann Arbor, MI 48109, USA \\ rischbie@umich.edu}

\maketitle
\begin{abstract}

\section{}

This paper will discuss the microphysical simulation of interactions in liquid xenon, the active detector medium in many leading rare-event searches for new physics, and describe experimental observables useful for understanding detector performance. The scintillation and ionization yield distributions for signal and background will be presented using the Noble Element Simulation Technique (NEST), which is a toolkit based on experimental data and simple, empirical formulae, which mimic previous microphysics modeling, but are guided by data. The NEST models for light and charge production as a function of the particle type, energy, and electric field will be reviewed, as well as models for energy resolution and final pulse areas. NEST will be compared to other models or sets of models, and vetted against real data, with several specific examples pulled from XENON, ZEPLIN, LUX, LZ, PandaX, and table-top experiments used for calibrations.

\tiny
 \keyFont{ \section{Keywords:} WIMP dark matter direct detection, liquid Xenon, simulations/models}
\end{abstract}
\vspace{-17pt}
\section{Introduction}

For the past 15+ years, leading results in dark matter direct detection searches have come from detectors based on the principle of the dual-phase TPC (Time Projection Chamber) using a liquefied noble element as the detection medium~\cite{Baudis:2018bvr}. Liquid xenon (LXe) TPCs have in particular produced the most stringent cross-section constraints, for spin-independent (SI) as well as neutron spin-dependent (SD) interactions between WIMPs (Weakly Interacting Massive Particles) and xenon nuclei. More recently, usage of LXe has also led to WIMP limits using different EFT (Effective Field Theory) operators, for mass-energies above $O$(5~GeV)~\cite{Akerib_2021_LUXRun03EFT}. EFTs extend the set of allowable operators beyond the standard SI and SD interactions, and include searches at higher nuclear recoil energies. Unrelated to dark matter, electron-recoil searches up to the MeV regime have set strict constraints on $0\nu\beta\beta$ decay~\cite{Anton:2019wmi}, and led to seeing double $e^-$ capture~\cite{Aprile:2019dec}. XENONnT and PandaX have recently illustrated the potential for precision measurements of $^8$B~\cite{aprile2024measurementsolar8bneutrinos,pandaxcollaboration2024indicationsolar8bneutrino}.

To interpret results from past, present, and future experiments, a reliable MC (Monte Carlo) simulation is required. Recent works have demonstrated the utility of NEST, the cross-disciplinary, detector-agnostic MC software reviewed here~\cite{Akerib_2021_LZSim,Yan_2021,Aprile_2021_8B}, for a variety of active detector materials: LAr~\cite{TokenDUNE,PhysRevD.107.092012,westerdale2023deap3600} and GXe, but especially LXe. As the multi-tonne-scale TPCs have commenced data collection~\cite{Aalbers_2023_LZsr1,Yan_2021,Aprile_2021_8B}, improved MC techniques will not only assist in limit setting, but will be needed for extracting dark matter particle mass and cross section in the event of a WIMP discovery. In either scenario, or for the design of a new TPC, predictions of performance are needed on key metrics like the fundamental scintillation light and ionization charge yields for LXe, the focus of this work. NEST v2.4 is its default; different versions are specified as needed. This manuscript is a technical overview of updates to NEST, including new models and comparisons. More pedagogical reviews of the models and related physics are available in \cite{NEST2011,instruments5010013}.

Section 2.1 contains mean scintillation and ionization yields of electronic recoil (ER) backgrounds, with comparisons to data. These underlie the ER background (BG) models in Xe-based dark matter detectors. Section 2.2 summarizes the methods for varying these mean yields to model realistic fluctuations, with variation in the total number of quanta (light and charge) produced. Section 2.3 has the yields of nuclear recoils (NR) and their fluctuations. These form the foundation for the signal model in a LXe-based dark matter search, as well as for NR backgrounds (from fast neutron scattering and coherent elastic neutrino-nucleus scattering, CE$\nu$NS). Lastly, Section 3 compares NEST's modeling of mean yields (Sections 2.1 and 2.3) to past and present approaches in existing literature, including some first-principles ones, before the conclusion. The strengths and weaknesses of the different approaches will be summarized, underscoring NEST's ability to model data across a broad range of energies and electric fields phenomenologically.

\begin{figure*}[th!]
\centering
\includegraphics[width=1\textwidth,trim=3.5cm 1cm 3.7cm 2cm]{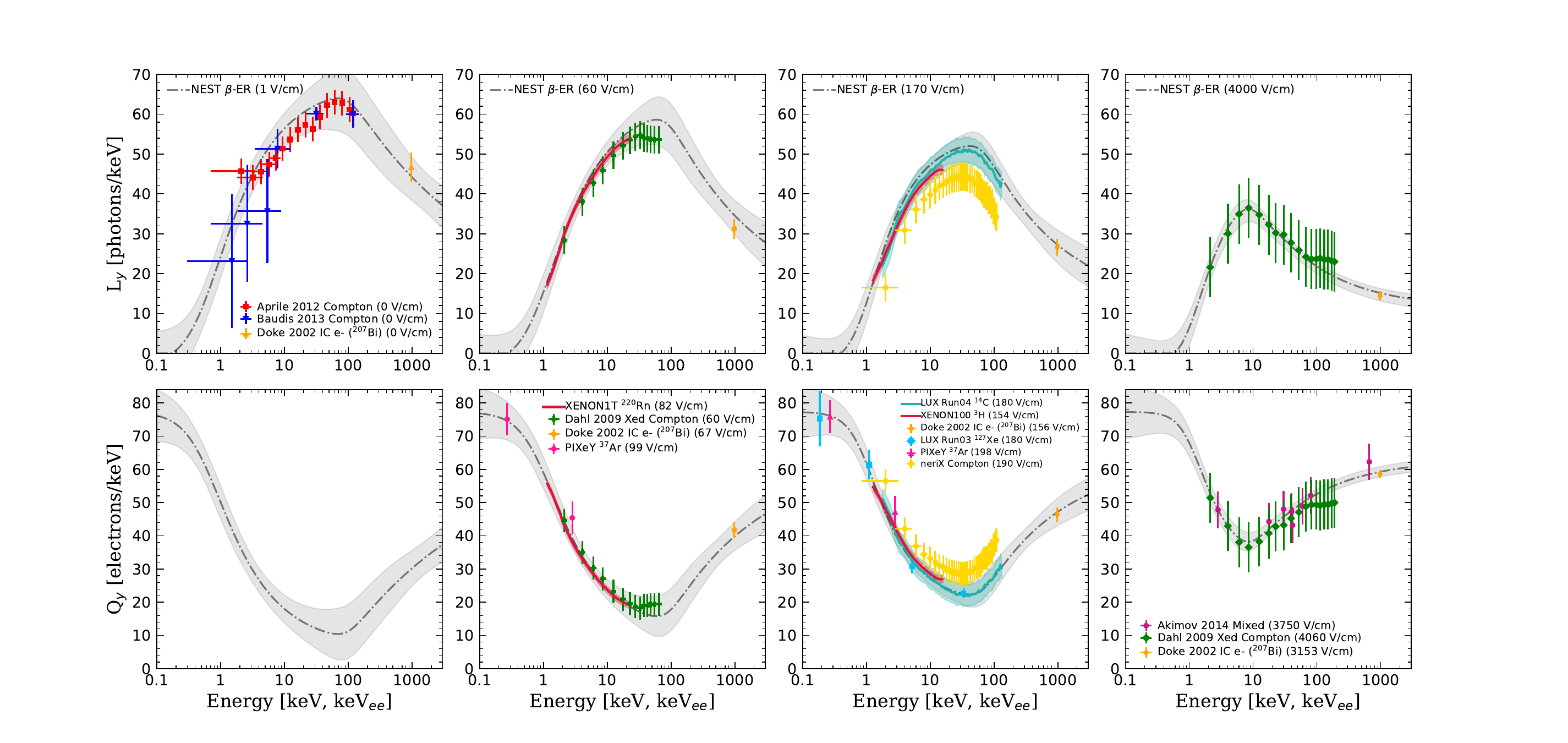}
\caption{$\beta$ electron recoil (ER) $L_y$ (top row) and $Q_y$ (bottom row) vs.~energy $E$. Different fields $\mathcal{E}$ are represented, from 0~V/cm (left column) to the highest fields for which data exist at multiple $E$s, $\sim$3--4~kV/cm (right column). More data exist, all of which are utilized to inform NEST, but these are selected as representative examples of the lowest and highest $\mathcal{E}$s and lowest and highest $E$s, from sub-keV to 1~MeV across different types of experiments~\cite{Aprile_2012_Compton,Baudis_2013,Doke_2002,Aprile:2019dme,Dahl:2009nta,pixey,Akerib:2019jtm,Aprile_2018_Discrim,Akerib_2017_DQ,Goetzke:2016lfg,Akimov_2014}. MC lines are black dashed with gray 1$\sigma$ error bands. Newer results \textit{e.g.} XENON1T's $^{220}$Rn calibration illustrate the predictive power of NEST, using the latest $\beta$ model which stems largely from $^{14}$C decays~\cite{Akerib:2019jtm,GregRC14}.}
\label{Fig1}
\end{figure*}

\section{\label{sec:second}Microphysics Modeling Evaluation}

NEST model choices were justified earlier (in \cite{instruments5010013} and in references therein) but are re-evaluated herein more comprehensively, with more, newer data. NEST is openly shared, allowing for regular re-evaluation using the latest calibrations~\cite{NESTWebSite}. While such data often provide relative light and charge yields, these are converted to absolute yields if the detector gains are calculable, known as $g_1$ and $g_2$ for those respective yields. The light yield gain, $g_1$, is the primary photon detection efficiency, while the charge gain, $g_2$, is the average signal size per $e^-$ escaping the interaction site. Uncertainties in these gains are a significant source of systematic uncertainty, but newer data from higher-quality calibrations mitigate this. Combining calibration data ranging from $<$ 1~keV to $>$ 1 MeV energy, NEST has predictive shapes for the primary scintillation and ionization yields as functions of energy, $E$, and drift electric field, $\mathcal{E}$, for different particle interaction types~\cite{Conti_2003}. The status of the NEST modeling of these shapes is shown in Figure~\ref{Fig1}.

\subsection{Electronic Recoils (Betas, Gammas, X Rays)}

NEST begins with a model of the sum total yield, summing the VUV (vacuum ultraviolet) scintillation photons and ionization electrons produced. IR photons are not included, as their yield in LXe is lower by a factor of $\sim$4~\cite{BRESSI2001378}, and their wavelength is beyond the sensitivity of most photon sensors in common use in dark matter experiments at least. The work function, $W_q$, for production of quanta depends only upon the density, using a linear fit based on data collected in \cite{Aprile:2008bga}, across phases (see also Appendix~A):
\vspace{-10pt}
\begin{equation}
W_q~[eV] = 21.94 - 2.93 \rho = \frac{W_i}{ 1 + N_{ex}/N_i }.% ~\mathrm{W_i,~often~confused~with~W_q,~was~defined~for~N_e-}.
\label{eqn:1}
\end{equation}

\noindent
$\rho$ is the mass density in units of g/cm$^3$. LXe TPCs  typically operate at temperatures of 165--180~K, and pressures of 1.5--2~bar(a), leading to $\rho \approx 2.9$~g/cm$^3$ and resulting in a $W_q$ of between 13--14~eV~(Eqn.~\ref{eqn:1}, with discrepant values discussed in \cite{NESTExcess}). The exciton-ion ratio or $N_{ex}/N_i$ relates $W_q$ with the work function for ionization, $W_i$, which was defined for charge yields. Moreover, $N_{ex}/N_i$ determines a pre-recombination (ions and $e^-$s) split of quanta into light and charge (see Appendix~A, where the $\rho$ dependence is explained):

\vspace{-15pt}
\begin{equation}
 N_{ex}/N_{i} = (0.0674 + 0.0397 \rho) \times \mathrm{erf}(0.05 E),
\label{eqn:2}
\vspace{-3pt}
\end{equation}

\noindent
where $E$ is deposited energy in keV, for a $\beta$ interaction or Compton scatter, and `erf' refers to the error function. Here, the $\rho$ dependence is based again on \cite{Aprile:2008bga} while the $E$ dependence comes from reconciling \cite{Doke_2002,Akerib:2015:tritium,Lin:2015jta}, given evidence that light yield approaches zero as energy $E$ decreases, with lower-$E$ data sets favoring both less recombination and smaller  $N_{ex}/N_{i}$. Ionization electrons can recombine with Xe atoms or can escape from the interaction site entirely. Therefore, the number of photons $N_{ph}$ is not simply equal to $N_{ex}$, providing an anti-correlation in the observed light and charge yields, motivating the use of both charge and light to measure the energy, $E = W_q~(N_{ph} + N_{e^-})$~\cite{instruments5010013}:

\vspace{-15pt}
\begin{equation}
\begin{split}
N_{ph} = [N_{ex} + r(E,\mathcal{E},\rho)N_{i}] = \mathrm{S1} ~/~g_1~~\mathrm{and}~~
N_{e^-} = [1 - r(E,\mathcal{E},\rho)]N_{i} = \mathrm{S2}~/~g_2,
\end{split}
\label{eqn:3}
\vspace{-3pt}
\end{equation}

\noindent
where $r$ is recombination probability for $e^{-}$-ion pairs depending on $E$, $\mathcal{E}$, and $\rho$, as well as particle and interaction type, and the S1 and S2 are the experimental observables. Typical values of $g_1$ are $\sim$0.1, but $O$(10) for $g_2$ due to secondary (gas) scintillation ($g_2$ is 0.5--1 in single-phase TPCs). The light and charge yields per unit energy are traditionally quoted in experiment, defined as $L_y~\equiv~N_{ph}/E$ and $Q_y~\equiv~N_{e^-}/E$.

$Q_y$ is modeled first; $L_y$ is set by $W_q$ and subtraction:

\vspace{-15pt}
\begin{flalign}
\begin{aligned}
N_q \equiv N_{ex} + N_i = N_{ph} + N_{e^-} = E~/~W_q,~~\mathrm{where}~~
N_{e^-} = Q_y E,~~\mathrm{and}~~
N_{ph} = N_q  - N_{e^-},
\end{aligned}
\label{eqn:4}
\vspace{-3pt}
\end{flalign}

\noindent
where $N_q$ is the total quanta. This procedure leverages the greater reliability of S2 measurements \textit{cf.}~S1 for lower $E$, as explained in \cite{Akerib_2017_DQ,instruments5010013}. $Q_y$ in the ER model is a sum of two sigmoids:

\vspace{-14pt}
\begin{equation}
Q_y ( E, \mathcal{E} ) =
m_1(\mathcal{E}) + \frac{m_2 - m_1(\mathcal{E})}{[ 1 + (\frac{E}{m_3(\mathcal{E})})^{m_4(\mathcal{E})}]^{m_9}} 
+ m_5(\mathcal{E}) - \frac{m_5(\mathcal{E})}{[ 1 + (\frac{E}{m_7(\mathcal{E})})^{m_8}]^{m_{10}(\mathcal{E})}},
\label{eqn:5}
\vspace{-2pt}
\end{equation}

\noindent
with $m_1$ serving as the minimum field-dependent charge yield. $m_2$ determines the low-$E$ behavior, and $m_7$ controls the field dependence at high energies. The individual $m_i$ are summarized within Appendix~B (with \cite{GregRC14} providing more details). While empirical, the first (left, $m_1$+...) and second (right, +$m_5$...) sigmoids of Equation~\ref{eqn:5} capture the qualitative behavior of two first-principles options, respectively: the Thomas-Imel box model at low energies~\cite{ThomasAndImel}, and Doke-modified Birks' Law at higher energies~\cite{Doke_1988}. Between $\sim$15~keV and the energy of a MIP (minimally ionizing particle) within Xe (about 1 MeV), a track shape is described as cylindrical by Doke for modeling the recombination, and the $dE/dx$ decreases with increasing $E$. The recombination probability $r$ decreases as energy $E$ increases, reducing the ratio of $L_y$ to $Q_y$~\cite{szydagis_m_2022_6989015,NEST2011,ESTARWebSite}. Below $\sim$15~keV, deposits are more amorphous, and straight 1-D track lengths become ill-defined: $r$ and $L_y$ increase with the 3-D ionization density and the energy, as $dE/dx$ increases with $E$.

A Thomas-Imel approach historically uses $E$ and models energy deposits within symmetric boxes or spheres, while the Doke~/~Birks' Law uses $dE/dx$ and assumes long tracks (cylinders). The former will exhibit $r$ (and therefore $L_y$) only increasing with energy, the latter decreasing usually ($Q_y$ anti-correlated).

\vspace{0.05in}
\begin{figure*}[th]
\centering
\includegraphics[width=1\textwidth,trim=3.5cm 1cm 3.7cm 2cm]{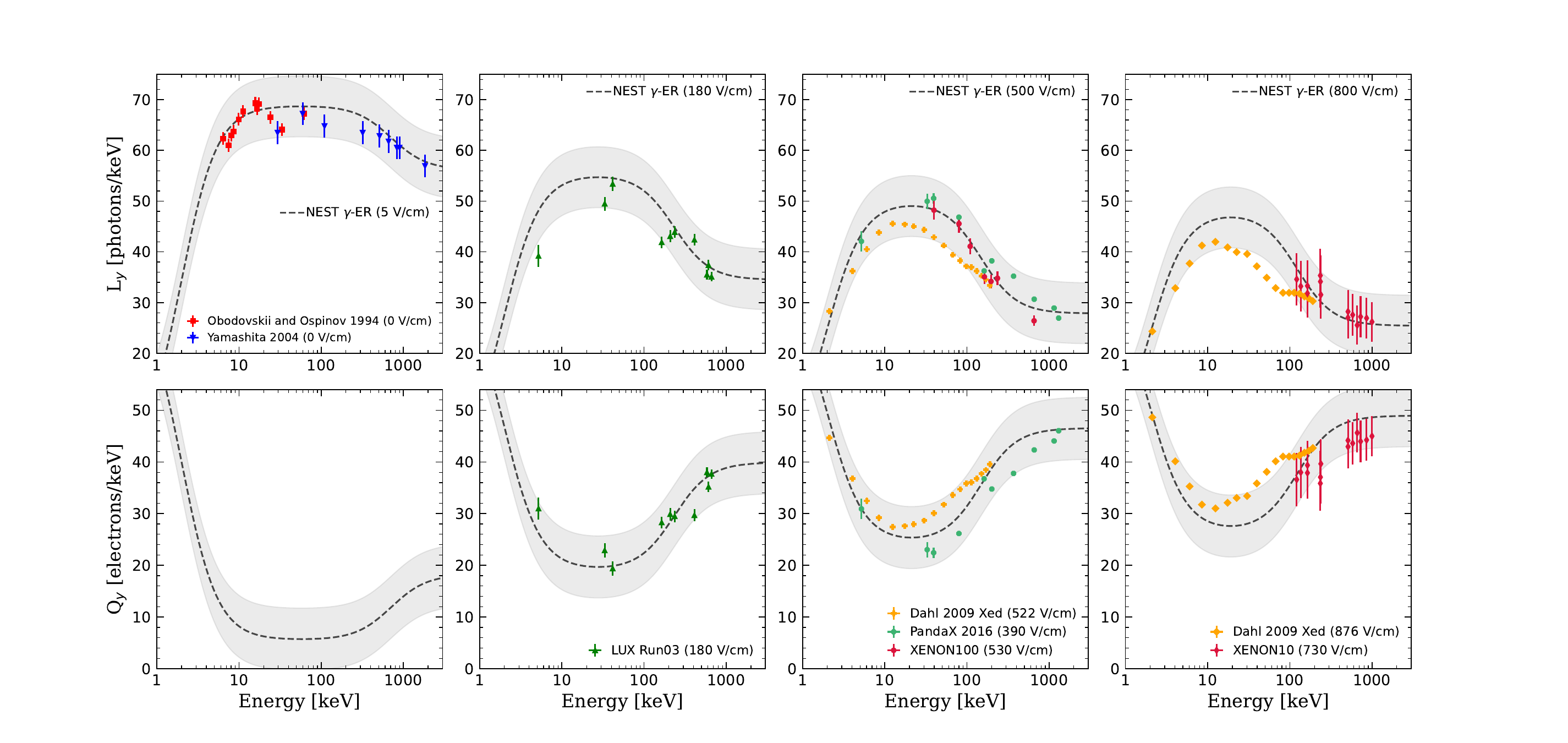}
\vspace{-12pt}
\caption{$\gamma$ ER $L_y$ (top row) and $Q_y$ (bottom) vs.~$E$ at $\mathcal{E}=0$ (left) to nearly 10$^3$~V/cm (right). Before $\beta$ calibrations were common, photoabsorption peaks from monoenergetic $\gamma$s were used~\cite{OboAndOsp_1994,YAMASHITA2004692,Akerib_2017_EvanP,Dahl:2009nta,PhysRevLett.117.121303,bib:50,bib:37}. At sufficiently high $E$, $L_y$ is higher and $Q_y$ lower than in Fig.~\ref{Fig1}, as some unresolvable multiple scattering occurs, treated as single scattering in NEST~\cite{Szydagis_2013}. Multiple lower-$E$, higher-$dE/dx$ vertices are ``averaged over.''  Low fields again approximate 0~V/cm, when NEST becomes singular. As in other plots gray 1$\sigma$ bands are driven by data errors, model shape constraints (sigmoidal), and monotonic $\mathcal{E}$ dependence. LUX $L_y$ points, but not $Q_y$, seem systematically low due to a different $W_q$ applied, with LUX assuming 13.7~eV (no $\rho$ dependence). Dahl data sets exhibit different shapes due to being mixtures of Compton scatters and photoabsorption.}
\label{Fig2}
\end{figure*}
\vspace{-10pt}
The recombination fraction or probability, $r$, is found retroactively in recent NEST versions after fitting to $Q_y$ per Equation~\ref{eqn:5}, chosen for matching both the box and Birks models. Using Equation~\ref{eqn:2} as a constraint avoids the degeneracy of this $r$ with $N_{ex}/N_i$, with the sum $N_{ex}+N_i$ (also equal to $N_{ph}+N_{e^-}$) already constrained by Equations~\ref{eqn:1} and \ref{eqn:4} -- the former determines $W_q$ and the latter total quanta $N_q$ based on $W_q$. Any change in $W_q$ (one work function averaging over individual work functions for photon and electron production) should change $L_y$ and $Q_y$ equally, preserving both their shapes in both energy and field~\cite{Anton:2019hnw}.

Figure~\ref{Fig1} summarizes both $L_y$ and $Q_y$ for $\beta$s and for Compton scattering ERs from both data and NEST, with NEST using a typical LXe operating condition of $\rho=2.89$~g/cm$^3$ ($T$ = 173~K, $P$ = 1.57~bar). The non-monotonic energy dependence is obvious. Meanwhile, the $L_y$ decreases from left to right (top) and $Q_y$ correspondingly increases (bottom) as field increases, suppressing recombination at a fixed $N_q$. But even at $\mathcal{E}=0$ there exists a ``phantom'' $Q_y$ likely caused by an extreme delay in recombination, as explained in \cite{Doke_2002, instruments5010013}, and repeated here: this is unobservable, except via long S1 integration times and by noting $L_y$ vs.~energy is the same shape at all fields, even 0. That implies a continuous change in $L_y$ as $\mathcal{E}\to0$. Non-0 fields standing in for 0 represent residual stray fields in a detector and/or inherent fields of Xe atoms~\cite{Szydagis_2013}.

Absorption of any high-energy photon, a $\gamma$ or x-ray, is modeled like $\beta$ interactions and Compton scatters, but with unique $m_i$ (Figure~\ref{Fig2}), to capture sub-position-resolution multiple scatters and distinct $dE/dx$. $L_y$ is mostly lower and $Q_y$ higher for $\beta$s, as explained within the Figure~\ref{Fig2} caption. While it might be possible to merge the $\gamma$ and $\beta$ models by relying on differences in $dE/dx$, $\gamma$s are treated independently at present. Appendix~B lists the $\beta$ and $\gamma$ model parameters, in addition to those for NR models.

\subsection{\label{sec:YieldFluctuations}Yield Fluctuations}
\vspace{-4pt}
Energy resolution typically refers to Gaussian spreads ($\sigma$ or FWHM) of monoenergetic peaks from high-energy $\gamma$-ray photoabsorption, but this is also relevant for lower energies, in WIMP searches. Smearing of continuous ER spectra can drive an increase in signal-like background events. But to understand statistical limitations for high-level parameters like monoenergetic-peak $\sigma$s or background discrimination we must start with lower-level parameters behind all the relevant stochastic processes involved. This modeling is discussed in depth in \cite{instruments5010013}, but portions germane to this work are summarized in this section, culminating in a subsection enumerating the practical steps taken within the NEST code on git.
\vspace{-4pt}
\subsubsection{Total Quanta: Correlated Fluctuations}
\vspace{-4pt}
Realistic smearing of mean yields begins with a Fano-like factor, $F_q$, applied to the total quanta, $N_q$, prior to differentiation into $N_{ex}$ and $N_i$. It is labeled as Fano-like, as it does not follow the strict sub-Poissonian definition~\cite{DOKE1976353}. $F_q$ may exceed 1, but it is still used in the usual definition of the standard deviation of $N_q$, utilized for decades by Xe experiments to fit their data on combined-$E$ ($N_{ph}$ and $N_{e-}$) scale resolution:
\vspace{-5pt}
\begin{equation}
\sigma_q=\sqrt{F_q \langle N_q \rangle},
\label{eqn:6}
\end{equation}
\vspace{-9pt}
\noindent
where $F_q$ is defined for light and charge together as
\vspace{-5pt}
\begin{equation}
F_q = 0.13-0.030\rho-0.0057\rho^2+0.0016\rho^3  + \delta_F \sqrt{\langle N_q \rangle} \sqrt{\mathcal{E}} . %~\mathrm{The~()~term~applies~only~to~LXe}.
\label{eqn:7}
\vspace{-2pt}
\end{equation}

\noindent
The first part of Equation~\ref{eqn:7} is a spline of data~\cite{Aprile:2008bga} from gas, liquid, and solid. The constant 0.13 represents the theoretical value of the Xe Fano factor following the traditional definition ($F_q<1$). $O$(0.1) matches NEXT gas data on $N_{e-}$~\cite{Alvarez:2012kua} and Biagi's Degrad work. The second part of Eqn.~\ref{eqn:7} is only for liquid and is data-driven, where $\delta_F=0.0015$ for LXe but is identically 0 for gaseous Xe. The $\sqrt{\langle N_q \rangle}$ term is included in order to match the data at MeV scales (\textit{e.g.}, for $0\nu\beta\beta$ searches). Such results did not achieve the theoretical minimum in energy resolution even when reconstructing $N_q$, utilizing both channels of information (light and charge), instead of only a single channel. This was true even for the cases where the noise was allegedly subtracted or modeled~\cite{Delaquis_2018,Aprile_2020_resE}. As $Q_y$ increases with $\mathcal{E}$, the combined $E$ resolution improves. However, the improvement is smaller than na\"ively predicted, requiring the $\sqrt{\mathcal{E}}$ term in $F_q$ to match the data~\cite{Aprile_2007,APRILE1991177}.

There are many possible explanations for this $F_q$ becoming $\gg$ 1 as $E$ or $\mathcal{E}$ changes. $W_q$ may need to be replaced with separate $W_{ex}$ and $W_i$ for the excitation and ionization processes (both inelastic scattering), then further subdivided into different values that depend upon $e^-$ energy shell. Lastly, elastic scattering of orbital $e^-$s may play a role. Mechanisms are discussed in \cite{platzman} but explicit Fano-factor variations can be found in \cite{instruments5010013}. In NEST, a Gaussian smearing, constrained to be non-negative, is applied to $N_q$ of width defined by Eqn.~\ref{eqn:6}: $N_q=G[\langle N_q \rangle,\sigma_q]$. A binomial distribution then divides quanta into excitons versus ions.
\vspace{-18pt}
\subsubsection{Anti-Correlated Excitation and Recombination Fluctuations}
\vspace{-4pt}
$F_q$ drives resolution on a combined-$E$ scale, but such a scale is more relevant for monoenergetic peaks than dark matter searches~\cite{Dahl:2009nta,instruments5010013}. ``Recombination fluctuations,'' however, describe the redistribution of $N_{ph}$ and $N_{e-}$ caused by widths paired with the means of Equation~\ref{eqn:3} or \ref{eqn:4}. Often conflated with excitation fluctuations (Equation~\ref{eqn:2}), these are all fundamental and do not originate from detector effects~\cite{bib:37,Akerib_2017_EvanP}, and constitute one of the key factors for characterization of ER discrimination~\cite{Dobi:2014wza}.  Moreover, they are not binomial, despite recombination (or, escape) appearing to be a binary decision. Potential explanations for this phenomenon include other energy loss mechanisms, or other effects which break the independence of draws, for instance $\delta$-ray production (as observed at different energies in both Ar and Xe~\cite{AMORUSO2004275,ThomImelBiller}), the statistics of columnar recombination~\cite{Nygren_2013}, and short-lived clustering of Xe dimers~\cite{Davis:2016reu}.

While it is unclear which explanation is correct, NEST proceeds with a fully empirical approach to simply model what is observed in data; following \cite{Akerib_2017_EvanP,GregRC14} closely, NEST defines recombination variance as:
\vspace{-5pt}
\begin{equation}
\begin{split}
\sigma_r^2 = \langle r \rangle ( 1 - \langle r \rangle ) N_i + \sigma_p^2 N_i^2,~~\mathrm{where}~~
\sigma_r \approx \sigma_{N_{e^-}} \approx \sigma_{N_{ph}} ~\mathrm{(for~small}~F_q), \\
\sigma_p = A(\mathcal{E}) e^{\frac{-(\langle y \rangle-\xi)^{2}}{2\omega^{2}}} [ 1 + \mathrm{erf} ( \alpha_p \frac{\langle y \rangle-\xi}{\omega \sqrt{2}} ) ],~
\mathrm{and~the~e^-~fraction}~y = N_{e^-} / N_q~\mathrm{and}~\langle y \rangle = Q_y W_q.
\label{eqn:8}
\end{split}
\end{equation}
\vspace{-3pt}
\noindent
The $\langle r \rangle (1-\langle r \rangle)N_i$ in $\sigma_r$ follows the binomial expectation of $\sigma_r \propto \sqrt{N_i}$. The $\sigma_p$ term leads to $\sigma_r \propto N_i$, as proposed in \cite{Dobi:2014wza}. $\sigma_p$ is a skewed Gaussian (on the second line) with field-dependent amplitude, $A$, varying from 0.05--0.1, as needed to simulate the spectral broadening of ER with higher drift electric field~\cite{GregRC14,VelanDiscrim}. In NEST versions $<$~2.1, $\sigma_p$ was simulated as a constant, similar in value to $A$, but this was found to be inadequate for capturing the full behavior of recombination fluctuations~\cite{Akerib_2017_EvanP}.

$\sigma_p$'s dependent variable was chosen to be the mean electron fraction $\langle y \rangle$ for simplicity, as it is closely related to 1$-\langle r \rangle$. Recombination probability, defined within Equation~\ref{eqn:3}, is degenerate with $N_{ex}/N_i$, while $y$ is directly measurable. It can be written in terms of $r$: $y=(1-r)/(1+N_{ex}/N_i)$~\cite{Dahl:2009nta}. Non-binomial fluctuations decrease as $y$ approaches 0 or 1, causing $\sigma_p$ to vanish. $\xi$, $\omega$, and $\alpha_p$ are the centroid, width, and skew of $\sigma_p$, respectively. Default NEST values determining respectively the width and skewness of $\sigma_p$ are $\omega=0.2$ and $\alpha_p=-0.2$. (Future work may recast all of $\sigma_r$ entirely in terms of $y$, not just $\sigma_p$.)

A skew centroid $\xi \approx$ 0.4--0.5 was found based on $\beta$ and $\gamma$ data. The types of data included continuous spectra and monoenergetic-peak energy resolutions, both at multiple fields and energies~\cite{Dahl:2009nta,bib:37,Dobi:2014wza}. $\xi$'s value depends on which data sets are used and which other parameters are fixed. A $\xi$ near 0.5 leads to a maximum in $\sigma_p$ (within $\sigma_r$) near $y=0.5$, as would occur within a regular binomial distribution. The asymmetric shape $\sigma_p$ is motivated by observations in recombination fluctuations at lower values of $y$ (low field, high energy) compared to higher values of $y$ (high field, low energy)~\cite{GregRThesis,Dobi:2014wza,GregRC14}.

Longer, less technical descriptions of all the steps within Section~2.2.2 are in \cite{GregRC14,GregRThesis}.

\vspace{-6pt}
\subsubsection{Recombination Skewness}
\vspace{-6pt}

We note here that the skewed Gaussian $\sigma_p(y)$ must not be conflated with the $E$ and $\mathcal{E}$-dependent skew defined in Section IVB of \cite{VelanDiscrim} as $\alpha_r$, as that skew is instead the observed asymmetry of the resultant charge yields. NEST uses $\alpha_r$ from Equation~(13) in~\cite{VelanDiscrim} to smear the mean $N_{e^-}$, while $\alpha_p$ controls the variance of recombination fluctuations, $\sigma_r$ from Equation~\ref{eqn:8}.

A positive $\alpha_r$ value can lead to better background discrimination than expected for a WIMP search that uses LXe. Weak rejection was expected due to the recombination fluctuations being greater (worse) than binomial, but positive $\alpha_r$ will shift ER events preferentially away from NR (more $Q_y$). This has already been observed~\cite{VelanDiscrim}.

\vspace{-6pt}
\subsubsection{Uncorrelated Fluctuations: Detector Effects (Known and Unknown)}
\vspace{-6pt}

\begin{figure*}[ht]
\centering
\includegraphics[width=1\textwidth,trim=3.5cm 1cm 3.7cm 2cm]{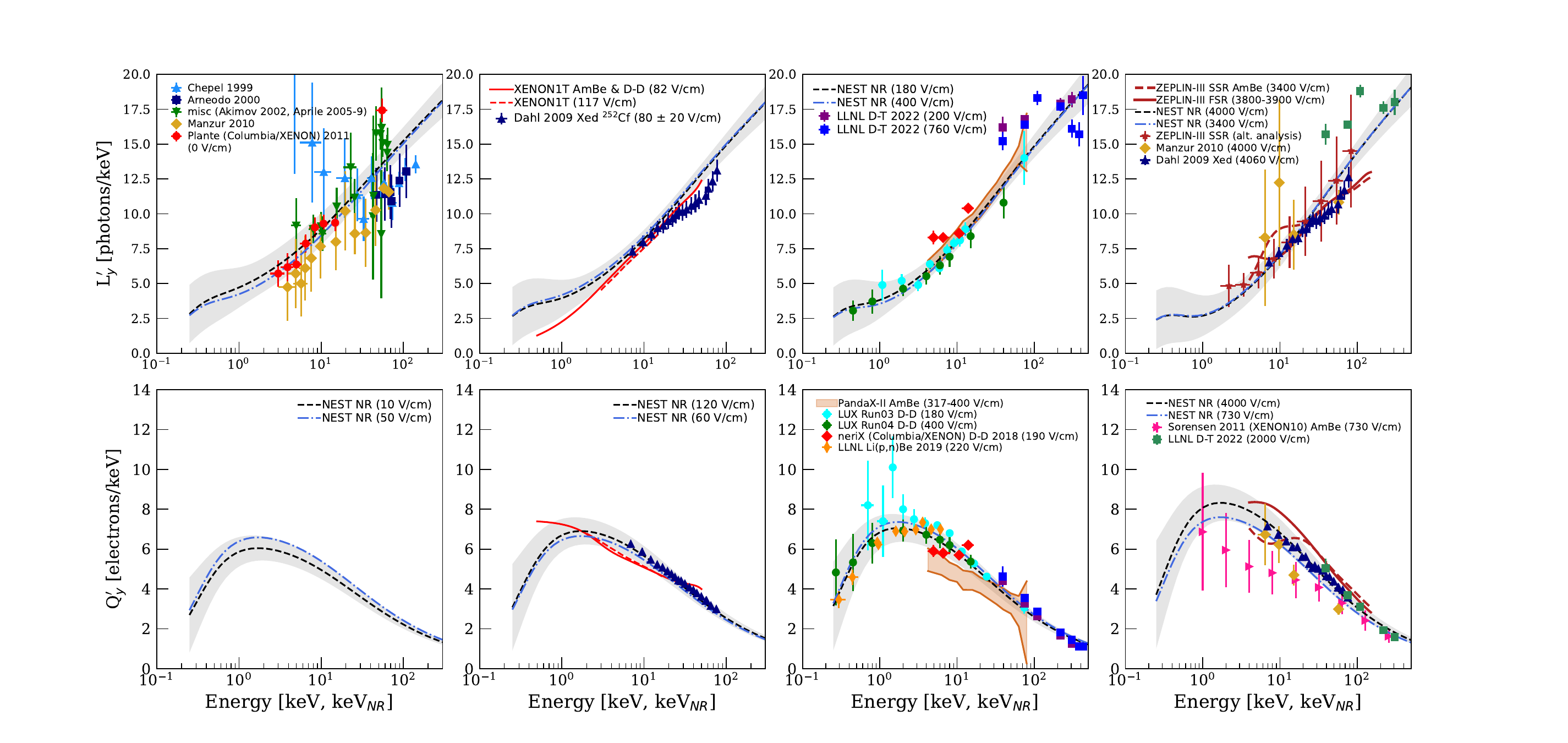}
\caption{(NR) $L_y'$ (top) and $Q_y'$ (bottom) vs.~$E$, from $\mathcal{E}=0$~V/cm at left to the highest $\mathcal{E}$ for which data exist at right~\cite{Aprile:2019dme,Dahl:2009nta,ChepelLeff1999,ArneodoLeff2000,AkimovLeff2002,AprileLeff2005,Aprile_2009,Manzur:2009hp,PlanteLeff2011,Aprile:2017iyp,Yan_2021,LUXDD2015,bib:LUX,Huang:2020ryt,DQLUXRun04DD,Aprile_2018_NeriXNR,lenardo2019measurement,Horn_2011,Sorensen:2010hv}. Newer works from XENON1T and PandaX were not included in fits (yet agree at the 1--2$\sigma$ level). NEST lines are blue and black, at similar $\mathcal{E}$s. Uncertainties on NEST increase as $E\to0$ or $\infty$, as the amount of data decreases at each extreme. $\mathcal{E}$ dependence is weaker compared with ER (Figure~\ref{Fig2}). Summing $L_y'$ and $Q_y'$ results in a power law, not a constant (ER), while $N_q'<N_q$~\cite{Sorensen_2011,instruments5010013}. For systematically-offset data sets, our fit can average them if they share the same qualitative trend. Discrepant results sharing the same trend point towards a systematic offset in the S1 and/or S2 gains, with the S1 most affected by the 2-PE effect~\cite{2PE} and the S2 affected by assuming 100\% $e^-$ extraction prior to more recent measurements~\cite{pixey2,llnl}. Only Chepel 1999 $L_y$ (upper left) is excluded from the fits used to tune NEST. As NR $dE/dx$ decreases with decreasing $E$, $e^-$ escape probability increases, causing $L_y'$ to decrease. ($L_y'$'s shape is also determined by the $\mathcal{L}$-factor.) For $Q_y'$, there is a maximum, as the $\mathcal{L}$-factor decreases and $(1-r)$ increases as $E \to 0$, at different rates. In contrast to \cite{instruments5010013}, where the focus was $\mathcal{L}$, we separate $L_y'$ and $Q_y'$ here. While errors imply no $\mathcal{E}$ dependence, when data are taken in one detector at many $\mathcal{E}$s, a rising $Q_y'$ (falling $L_y'$) with $\mathcal{E}$ is clear~\cite{Dahl:2009nta}.}
\label{Fig3}
\end{figure*}

Lastly, while the simulated $\sigma_q$ widths predict correlated changes in S1 ($L_y$) and S2 ($Q_y$), and $\sigma_r$ leads to anti-correlated change, uncorrelated noise also exists, affecting S1 and S2 independently. S1 and~S2 gains are understood sources, assuming position-dependent light collection and field non-uniformities are taken into account. Unknown sources are modeled with~a Gaussian smearing proportional to the pulse areas~\cite{NESTExcess}.~A quadratic term may be necessary at the MeV scale~\cite{Davis:2016reu}. ER and NR are equally affected by any detector effects (known/unknown). The final $E$ resolutions vs.~$E$ are seen for ER, NR, or both in \cite{Akerib_2021_LZSim,NESTExcess}, supplementing validation~of means in our Figures~\ref{Fig1}, \ref{Fig2}, and \ref{Fig3} with their vetting of fluctuations. The scale of the unknown detector effects across experiments is 1--10\%~\cite{NESTExcess,instruments5010013,lzcollaboration2023constraints} (for S2s, and for non-integer forms of S1s, but effectively 0\% for a spike count of S1 photons). See again Appendix~A.

\subsubsection{Computational Implementation}
\vspace{-2pt}
NEST is publicly available as a GitHub repository which includes the source code, interface scripts, and examples. It is C++ based, but can be run by dedicated scripts using either C++ or Python available on the repo. These can be used to generate expectation values of yields and their fluctuations for different detectors using Xe or Ar. The step-by-step procedure that NEST uses to do the latter is summarized below:

\vspace{-2pt}
\begin{itemize}
    \item $F_q$ is used to determine the $\sigma_q$ for a normal distribution of total (initially undifferentiated) ER quanta which can be considered ``correlated noise,'' because in this case $N_{ph}$ and $N_{e-}$ rise and fall together (Eq.~A1 \cite{xenoncollaboration2024xenonntwimpsearchsignal}). Two distinct $F$s exist for NR $N_{ex}$ and $N_i$, breaking that correlation (Section 2.3).
    \item ER quanta are differentiated ($N_{ex}$ and $N_i$) using a binomial distribution (Eq.~A2 \cite{xenoncollaboration2024xenonntwimpsearchsignal}), approximated as normal for computational speed, using the same Box-M\"{u}ller algorithm as in the first bullet above. Any non-binomial~/~non-Gaussian fluctuation at this stage is essentially degenerate with the next step.
    \item A normal or skew-normal (Eq.~8--12 in \cite{VelanDiscrim}) in $N_{e-}$ capped at $N_q$ (min of 0) enforces the anti-correlated fluctuation of $N_{ph}$ versus $N_{e-}$. This step was mismodeled in the past by uncorrelated Fano factors. The variance $\sigma_r^2$ has components proportional to both $N_i$ (``binomial style'') and $N_i^2$ (data-driven).
\end{itemize}
\vspace{-2pt}
Two more lists cover detector specifics for S1 and S2, closely following Appendix C of \cite{xenoncollaboration2024xenonntwimpsearchsignal}. First, S1:
\vspace{-2pt}
\begin{itemize}
    \item S1.1 A binomial distribution with probability $g_1$ (3-D spatially varying) determines the fraction of $N_{ph}$ successfully detected by photo-sensors; $g_1$ represents the product of geometric $\times$ quantum efficiencies.
    \item S1.2 Single photo-electrons in sensors are modeled by 0-truncated Gaussians of sensor-specific width. Spike counting is emulated by means of artificially reduced width but non-zero for matching real data.
    \item S1.3 An if-else structure determines whether a second photo-electron is produced due to the 2PE effect. This step and S1.2 are Gaussian-approximated at high $E$ in ``hybrid'' mode, or any $E$ in ``parametric.''
    \item S1.4 Geant4 (G4), Chroma, OptiX, or some other ray-tracer, or NEST's built-in analytic-approximation ability simulates photon arrival times at S1 sensors and dictates whether a sufficient number of photons were detected in MC with above-threshold (experiment DAQ-specific) pulse areas, based upon S1.2+3.
\end{itemize}
\vspace{-2pt}
The procedure to model the charge signal or S2 is more intricate, especially in a two-phase experiment:
\vspace{-2pt}
\begin{itemize}
    \item S2.1 Electrons (numbered $N_{e-}$) diffuse, transversely and longitudinally, as they drift at a drift speed determined by the liquid field but also density, and possibly temperature and pressure separately (the same applies to diffusion ``constants''). Data-driven functions exist for all these phenomena in NEST.
    \item S2.2 An electron survival fraction is set by an exponential function depending on the originating depth in a detector and a characteristic electron MFP. It is used as the probability in a binomial distribution.
    \item S2.3 Another binomial distribution is utilized to find how many electrons survive extraction from the liquid to the gas. The efficiency is a function of the gas field $\mathcal{E}_g$ between the liquid/gas boundary and the gate grid. NEST offers many options of asymptotic (1 at infinite $\mathcal{E}_g$) function based upon past data.
    \item S2.4 Each extracted electron produces $Y_{e-}$ S2 photons based on the parameterization in \cite{Chepel_2013} depending on $\mathcal{E}_g$, gas $\rho$, and the gap between the liquid surface and gate (thus $\mathcal{E}_g$ comes into play twice). $Y_{e-}$ is the mean of an int-rounded Gaussian of width $\sqrt{F_{S2}Y_{e-}}$. $F_{S2}$ is $O$(1) and captures grid non-uniformity.
    \item S2.5 A binomial of probability $g_1^{gas}$ (2-D varying) like $g_1$ (liquid) begins a process similar to S1.1--4.
\end{itemize}

\vspace{-2pt}
More precise S2 simulation is possible in the optional integration of Garfield with NEST, which also possesses an optional G4 integration for simulating $E$ deposits prior to the first step above. More details on the lists here can be found in Section 2.2 of \cite{instruments5010013}. Section 2.2.4 explains NEST's last layer. All values for the first list are in Table S4 (Appendix B) and examples for S1 and S2 in \cite{GregRThesis} especially in its Fig.~4.3 left.

\vspace{0.1in}

\subsection{Nuclear Recoils (Neutrons and WIMPs and Boron-8)}

NR $N_q'$ (differentiated in this section from ER with a prime) is well fit by a power law across $>$3 orders of magnitude in $E$ (Fig.~5 within \cite{instruments5010013}). This is a simplification of the Lindhard approach to modeling the reduced quanta compared with ER, but also allows for departures from Lindhard at higher $E$s, lowering $N_q'(E)$'s rate of change with respect to Lindhard. Fewer equations and parameters are involved compared to Lindhard, which is a combination of multiple power laws inside of a rational function~\cite{Lindhard_1963}: see Eqn.~8 in \cite{instruments5010013}, where more justification is provided. NEST uses that simpler formula, repeated here:

\begin{equation}
N_q' = a E ^ b, \mathrm{~where~} a = 11^{+2.0}_{-0.5} \mathrm{~and~} b = 1.1 \pm 0.05.
\label{eqn:9}
\end{equation}

\noindent
The uncertainties here are $>10\times$ those reported recently for the same fit, as only statistical error was included in Eqn.~6 of \cite{instruments5010013}. Here, systematic uncertainties in S1 detection efficiency and S2 gain (including $e^-$ extraction efficiency) are included. They can be found inside the individual references in the Figure~\ref{Fig3} caption. Individual power laws were found for each data set prior to the error-weighted combination, so that a data set with more points was not overly weighted. Equation~\ref{eqn:9} was also cross-checked with the $L_y'$ and $Q_y'$ individually extracted from data as displayed in Figure~\ref{Fig3}, and the raw S1 and S2 data of continuous energy spectrum sources.

Equation~\ref{eqn:9} can be used to define the ``quenching," $\mathcal{L}$:
\vspace{-1pt}
\begin{equation}
\mathcal{L}(E,\rho) = N_q'(E)~/~N_q(E,\rho) = N_q'(E)~W_q(\rho)~/~E,
\label{eqn:10}
\end{equation}
interpreted as the fraction of total NR energy shared with the electron cloud to produce ions and excitons. $\mathcal{L}$ permits one to define the electron equivalent energy in units of keV$_{ee}$ for NR, as $\mathcal{L} \times (E$ in keV$_{nr}$), a best average reconstruction of the (combined-)$E$ of recoiling nuclei. This $\mathcal{L}$ should be applicable to neutron calibrations, WIMPs, and CE$\nu$NS, such as from $^8$B nuclear fusion~\cite{Aprile_2021_8B}.

While the previous equation set total quanta, the next equation determines the field- and density-dependent division into individual yields (charge or light) in an anti-correlated fashion, reducing $r$ with higher field:

\begin{multline}
\varsigma(\mathcal{E},\rho) = \gamma \mathcal{E} ^{\delta} \left( \frac{\rho}{\rho_0} \right)^{\upsilon},~\mathrm{where~} \gamma = 0.0480 \pm 0.0021~\mathrm{and}~\delta = -0.0533 \pm 0.0068, \mathrm{~and~} \upsilon = 0.3.
\label{eqn:11}
\end{multline}

\noindent
The reference density $\rho_0 \equiv 2.90$~g/cm$^3$. (The 2.89 value earlier was a very specific example, using LUX; the differences in yields are negligible.) The exponent $\upsilon$ for the density dependence is hypothetical. It is not well measured at densities significantly deviating from $\rho_0$~\cite{Dahl:2009nta}.

We utilize Equation~\ref{eqn:11} to produce a $Q_y'$ equation:
\vspace{-1pt}
\begin{multline}
Q_y' ( E, \mathcal{E}, \rho ) = N_{e^-}~ \mathrm{per~keV} = \frac{1}{\varsigma(\mathcal{E},\rho)(E+\epsilon)^p} \left( 1 - \frac{1}{1+(\frac{E}{\zeta})^{\eta}} \right),  \mathrm{where} \\
\quad\quad \epsilon = 12.6^{+3.4}_{-2.9}~\mathrm{keV},~p=0.5, \zeta = 0.3 \pm 0.1~\mathrm{keV},~\mathrm{and}~\eta = 2 \pm 1.
\label{eqn:12}
\end{multline}

\noindent
Energy deposited is again $E$ (in keV), and $\epsilon$ is the reshaping parameter for the $E$ dependence. Higher or lower $\varsigma$ lowers or raises the $Q_y'$ level respectively, providing the field-dependent shape of $Q_y'(E)$. $\epsilon$ can be thought of as the characteristic $E$ where the $Q_y'$ changes in its behavior from $\sim$constant at $O$(1~keV) to falling at $O$(10~keV). (Note $\varsigma$ has adaptable units of keV$^{1-p}$.)

$\zeta$ and $\eta$ are the two sigmoid parameters that control the $Q_y'$ roll-off at sub-keV energies. They permit a better match to not only the most recent calibrations~\cite{lenardo2019measurement,bib:LUX}, but also to NEST versions pre-2.0, and other past models. 
Combining Thomas-Imel recombination with Lindhard (Eqn.~8 of \cite{instruments5010013}) produces a roll-off in $Q_y'$ which is less steep than observed in data. Here, $\eta$ controls steepness, allowing for an improved modeling of low-energy NR~\cite{Szydagis_2013,Sorensen_2011}, while $\zeta$ represents a characteristic scale for NR to ionize one $e^-$~\cite{instruments5010013,Sorensen_2015}. At high $E$, $p=0.5$ reproduces $Q_y' \propto 1/\sqrt{E}$ (Figure~\ref{Fig3}, bottom row).

Similar to ER, $N_{ph}$ is derived from $N_q'-N_{e-}$, but this is only a temporary anti-correlation enforcement;  an additional sigmoid permits $L_y'$'s flexibility (Equation~\ref{eqn:13}). Future calibration data could show a drop, or even flattening potentially, due to additional $N_{ph}$ from the Migdal effect~\cite{LUXDD2015,Aprile_2019_Light}. An $L_y'$ increase is possible even as $E \to 0$. This is not unphysical as long as $N_{ph}$ vanishes in that limit, conserving $E$.

\begin{multline}
L_y^{''} = \frac{N_q'}{E} - Q_y'.~N_{ph} = L_y^{''} E \left( 1 - \frac{1}{1+(\frac{E}{\theta})^{\iota}} \right);~
L_y' = \frac{N_{ph}}{E}, \\
\mathrm{where}~\theta = 0.3 \pm 0.05~\mathrm{keV}~\mathrm{and}~\iota = 2 \pm 0.5.~N_q' = N_{ph} + N_{e^-}.
\label{eqn:13}
\end{multline}

\noindent
The top row of Figure~\ref{Fig3}, especially if it is read from right to left, shows the same $L_y'$ shape at all fields, indicative once again of a zero-field phantom $Q_y'$. In the $L_y'$ calculation, $L_y^{''}$ is a temporary variable (perfect anti-correlation) used within NEST to calculate the final $L_y'$ and $N_q'$. The best-fit numbers for $\theta$ and $\iota$ match those of their counterparts $\zeta$ and $\eta$ for $Q_y'$. In this modular but smooth approach the sigmoidal terms in $L_y'$ and $Q_y'$ go to 1.0 with increasing $E$. In this fashion it is possible to fit the low- and high-$E$ regimes separately, allowing for a possibility that different physics occurs in the sub-keV region, by avoiding use of higher-$E$ data to over-constrain lower-$E$ yields.

The two sigmoids lower the predictive power of NEST for extrapolation into newer, lower-$E$ regimes where no calibrations exist. In the case of $L_y'$, it will be challenging to achieve any with low uncertainty.

$\theta$ is a physically-motivated characteristic energy for release of a single (VUV) photon. Like $\zeta$, its value is 300~eV, in agreement with Sorensen~\cite{Sorensen_2015}, and NEST pre-v2.0.0~\cite{Szydagis_2013}. Fundamental physics models for the $\mathcal{L}$ governing total quanta, such as Lindhard~\cite{Lindhard_1963} and Hitachi~\cite{Hitachi,PhysRevLett.97.081302}, coupled to the Thomas-Imel ``box'' model for recombination~\cite{ThomasAndImel}, predict a similar value. Larger $\theta$ means more $E$ is needed to produce a single photon (as opposed to excitons) and $L_y'$ is lowered. This may potentially be detectable for an experiment with sufficient light collection efficiency.

Decreasing $\iota$ would lower $L_y'$ as well, halving $L_y'$ across all $E$ when $\iota=0$. On the other hand, in the limit of infinite $\iota$ (and/or $\theta \to~0$) the effect of the sigmoid is entirely removed, raising $L_y'$ at low $E$. The same is true for $\eta$ and $\zeta$ in the $Q_y'$ formulation. A hard cut-off for any quanta was implemented in NEST for $E<W_q~(N_q~/~N_q') \approx 200$~eV. $N_q$ represents the quanta which would have been generated for same-$E$ ER. Below this, no quanta are generated. Sub-keV recoils have been observed at 200--400~V/cm (Figure~\ref{Fig3}).

In contrast to ER, for which the data suggest strict anti-correlation, simulated $\langle N_q' \rangle$ is not varied with a common Fano factor shared by both types of quanta for simplicity. For NR, there are (nominally) separate Fano factors for the excitation and ionization which can soften the strict anti-correlation at the level of the fundamental quanta. $\langle N_{ex} \rangle$ is smeared using a Gaussian of standard deviation $\sigma_{ex}$ = $\sqrt{F_{ex} \langle N_{ex} \rangle}$. $\langle N_i \rangle$ is similarly varied, using $\sigma_i=\sqrt{F_i \langle N_i \rangle}$, as is standard practice for Fano factors~\cite{PhysRev.72.26}. Based upon the sparse existing reports of NR $E$ resolution~\cite{LUXDD2015,lenardo2019measurement,Plante:2012umc} both $F_{ex}$ and $F_i$ are set to 0.4 in NEST (as of v2.3.11; 1 earlier) though some data imply $F_{ex}$ $\gg$ 1~\cite{LUXDD2015,Plante:2012umc}. $N_{ex}=G[\langle N_{ex} \rangle,\sigma_{ex}]$ and $N_{i}=G[\langle N_{i} \rangle,\sigma_{i}]$ ($G$=Gauss).

Using the same functional form as in Equation~\ref{eqn:8} from ER, NEST models fluctuations in recombination for redistribution of photons and electrons prior to measurable NR S1 and S2. The new parameters are distinguished using a prime symbol superscript again for NR ($'$).

Parameter values are similar but not identical to those from ER: $A'=0.04$ (as of v2.3.11 and fixed for all fields), $\xi'=0.50$, and $\omega'=0.19$ ($\alpha_p'$ = 0). Over time, these do appear to have been converging upon values similar to ER's. These set a final recombination width $\sigma_r'$. $N_{e-}$ and $N_{ph}$ distributions have that width but are skewed due to NR recombination asymmetry ($\alpha_r'$ = 2.25). $\alpha_r'$ may be higher, but it is difficult to disambiguate NR skew (less $L_y$) in data from unresolved multiple scatters, other detector effects~\cite{VelanDiscrim}, or Migdal-Effect ER, which can raise $Q_y$ and generate a secondary population~\cite{Akerib_2019_Migdal}.

\textcolor{white}{.}

\setcounter{figure}{4}
\setcounter{subfigure}{0}
\begin{subfigure}[ht!]
\setcounter{figure}{2}
\setcounter{subfigure}{0}
    \centering
    \begin{minipage}[b]{0.85\textwidth}
        \includegraphics[width=\linewidth]{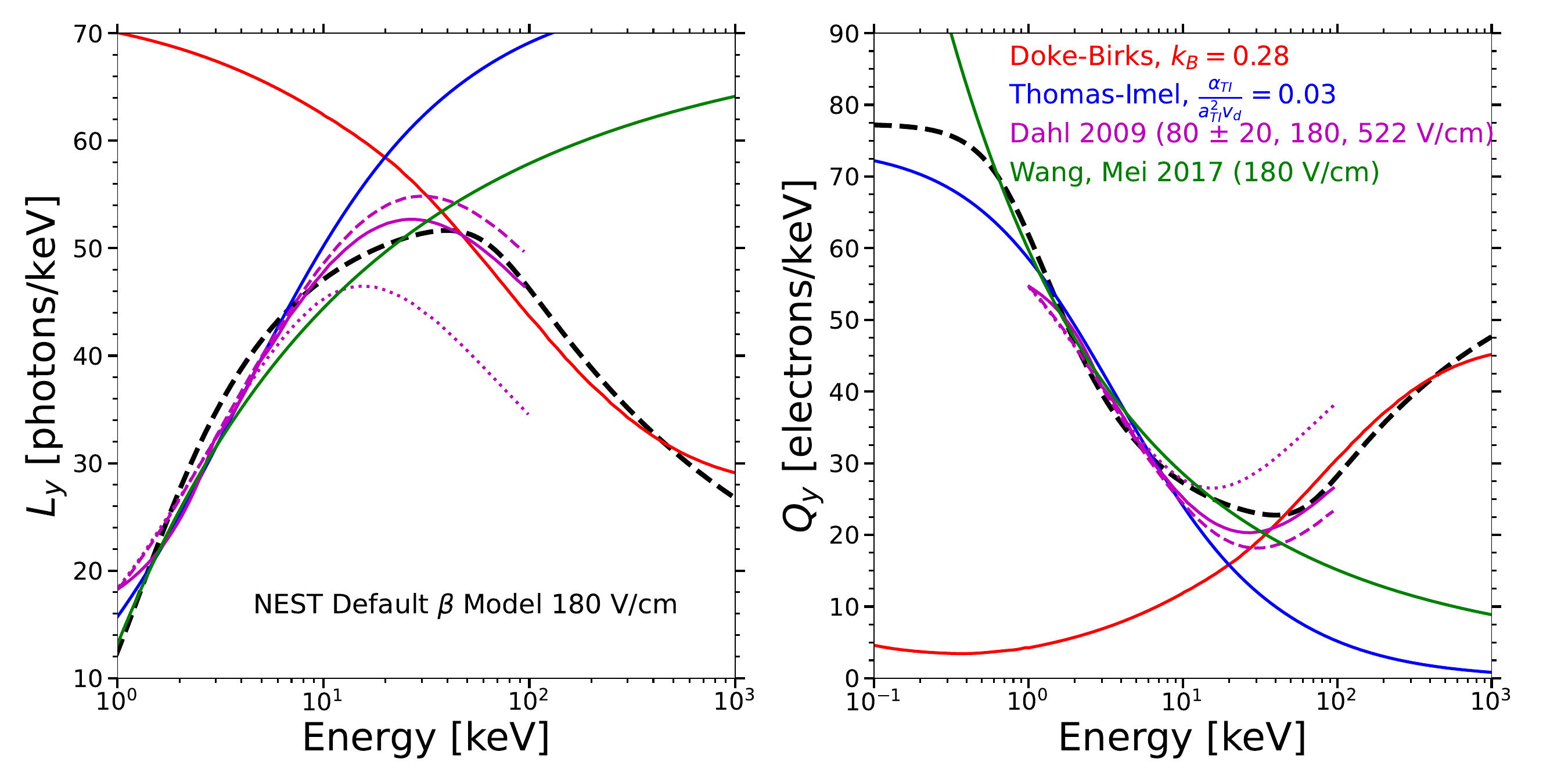}
        %\caption{This is Subfigure 1.}
        \label{Fig4a}
    \end{minipage}  
   
\setcounter{figure}{4}
\setcounter{subfigure}{1}
    \begin{minipage}[b]{0.85\textwidth}
        \includegraphics[width=\linewidth]{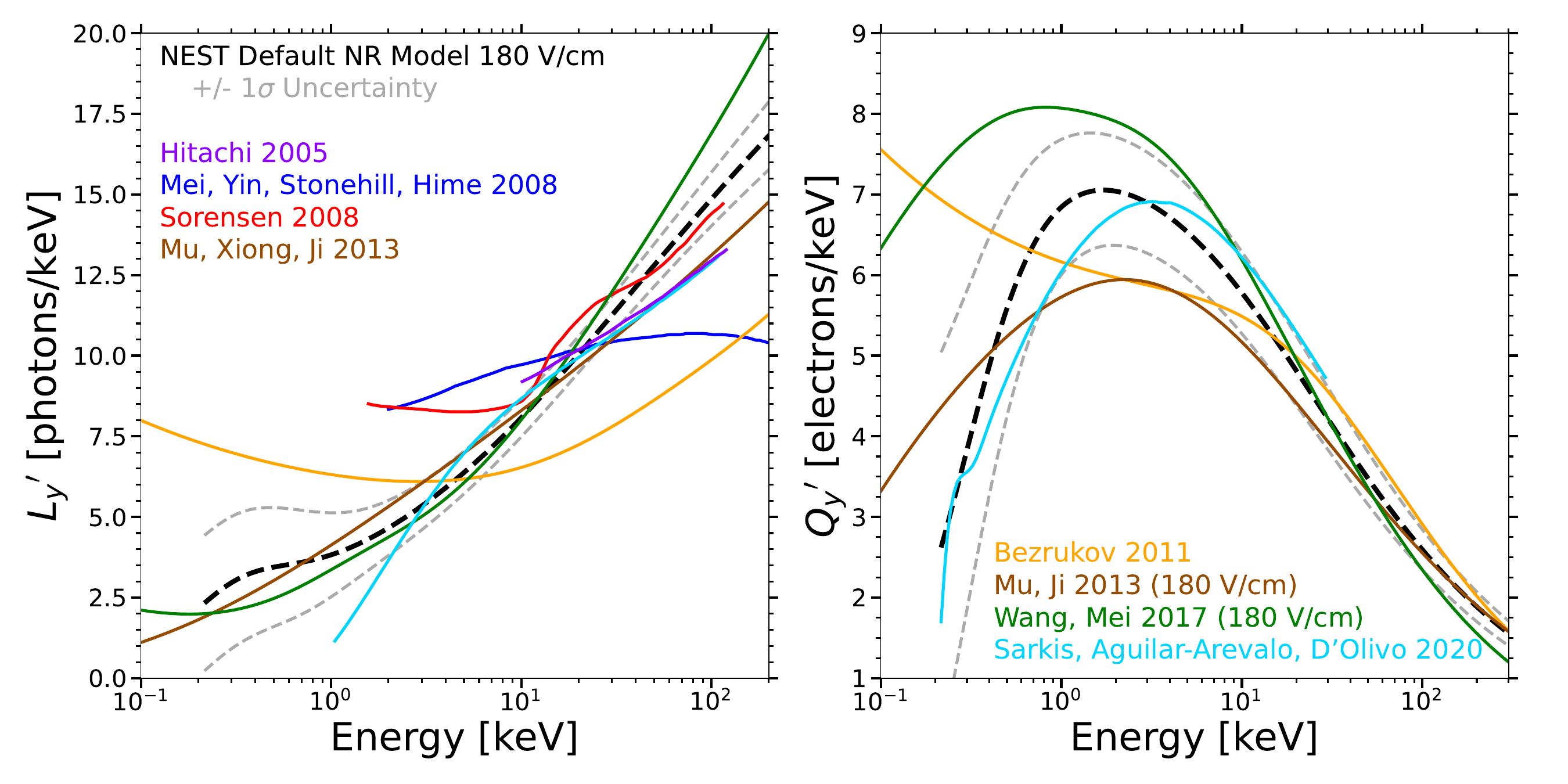}
        %\caption{This is Subfigure 2.}
        \label{Fig4b}
    \end{minipage}

\setcounter{figure}{4}
\setcounter{subfigure}{-1}
    \caption{Comparing NEST with other approaches: $L_y$ (left) and $Q_y$ (right) alternate, for ER (top) and NR (bottom), at 180~V/cm~\cite{Doke_2002,ThomasAndImel,Dahl:2009nta,Wang:2016obw}. The right legends apply to both the left and right plots. This was LUX's initial field~\cite{luxrun3_2013}, in between XENON1T at 80~\cite{Aprile_2020_Excess} and earlier works like \cite{bib:37} as high as 730~V/cm. While similar to fundamental approaches, NEST incorporates features of multiple, splitting differences and following the data. The Thomas-Imel (T-I) and Doke~/~Birks sample curves shown are meant to match 180~V/cm the most closely. Unlike the T-I and plasma models, NEST accounts for the high-$E$ (low-$dE/dx$) $L_y$ decrease ($Q_y$ increase)~\cite{Wang:2016obw}. Birks' Law does too, but fails to work at low $E$s (high $dE/dx$)~\cite{BirksTheoryAndPracticeOfScintillationCounting}. Dahl presented variations on T-I, utilizable for high $E$s by breaking up tracks into boxes, although his closest fields were 80 and 522~V/cm~\cite{Dahl:2009nta}. We show a 180~V/cm model (solid), \textit{i.e.}, the weighted average of his 80 (dashed) and 522~V/cm (dotted) models. There are more NR models (right), due to a need to explain potential WIMPs~\cite{Wang:2016obw,Hitachi,Mei_2008,Sorensen_2009,mu2013scintillation,Bezrukov_2011,mu2013ionization,Sarkis_2020}. Older models based on $L_{eff}$, which was $L_y'$ relative to $^{57}$Co $\gamma$-rays (122~keV), were translated assuming 64~photons/keV at 0~V/cm with a small error~\cite{NEST2011,NEST2015}, unless papers had a different value, which we then used instead (Bezrukov:~53). If they presented multiple models, we plot the most central one and/or one closest to data. Comparisons are only qualitative here, ensuring NEST has the correct, physically-motivated shape across different regimes.}
    \label{Fig4}
\end{subfigure}

\section{Comparisons to First-Principles Approaches}

By smoothly interpolating data sets taken at individual energies and/or electric fields NEST is now fully empirical, built upon sigmoids and power laws as needed for a continuous model. But inherent uncertainty is introduced by extrapolating into new energy and/or field regimes. To assess that, and to further validate an empirical approach, we show agreement with the models closer to ``first principles.'' Within NEST's earliest versions, the Thomas-Imel (T-I) box model~\cite{ThomasAndImel} was used for low energy, while Birks' Law of scintillation was adapted for high energy. Both were qualitatively explained in Section~2.1, but are quantified here. The latter approach inside NEST was similar to Doke's modification~\cite{NEST2011} for scintillation alone, but applied to recombination directly so it can model both the $L_y$ and the $Q_y$:
\vspace{-5pt}
\begin{equation}
\langle r \rangle = \frac{k_A \frac{dE}{dx}}{1+k_B \frac{dE}{dx}} + k_C,~\mathrm{with}~k_C = 1 - k_A / k_B.
\label{eqn:14}
\end{equation}

\noindent
This is Birks' Law for other scintillators~\cite{BirksTheoryAndPracticeOfScintillationCounting} but with an additional constant $k_C$ that accounts for parent-ion recombination~\cite{Doke_2002}. Its constraint ensures $\langle r \rangle$ is between 0--1, as it is a probability. A best fit to ER ($\gamma$) data has a non-zero $k_C$ only at 0~V/cm; at non-zero $\mathcal{E}$, Equation \ref{eqn:14} contains only one Birks' constant, $k_A = k_B$.

$k_B$'s best-fit value (for 180~V/cm) is 0.28, from a fit to only the high-$E$ portion of the NEST $\beta$ ER model. That is in turn supported by $^3$H, $^{14}$C, and ${}^{220}$Rn data from LUX and XENON. Notably, $k_B$ in NEST v0.9x and the first NEST paper 13 years ago for this $\mathcal{E}$ was 0.257, within 10\% of the value in Figure~\ref{Fig4} (upper right plot pane), which covers many alternative approaches to NEST.

Despite Birks' great success in explaining data at high $E$, that model cannot capture the behavior of ER at $E \lesssim$ 50~keV. While lower-$E$ extensions are possible, such as addition of higher-order terms in $dE/dx$ for that region, we instead consider the T-I model for lower $E$:
\vspace{-5pt}
\begin{equation}
\langle r \rangle = 1 - \frac { \mathrm{ln} ( 1 + \xi_{TI} ) } { \xi_{TI} },~\mathrm{where}~\xi_{TI} = \frac{N_i}{4} \frac{\alpha_{TI}}{a_{TI}^2 v_d}.
\label{eqn:15}
\end{equation}

\noindent
$\xi_{TI}$ parameterizes the physical principles. $\alpha_{TI}$ describes diffusion, $v_d$ is $e^-$ drift velocity, and $N_i$ is again number of ions. Diffusion is modeled by using the relation $\alpha_{TI}= D e^2 / ( k T \epsilon_d )$, where $D$ combines $e^-$ and positive-ion diffusion coefficients, $e$ is the elementary charge, $k$ is Boltzmann not Birks, $T$ is temperature, and $\epsilon_d= 1.85\times\epsilon_0$ is the dielectric constant. $D=18.3$~cm$^2$/s is the longitudinal diffusion constant for $e^-$s at 180~V/cm, derived from S2 pulse lengths~\cite{Sorensen_2011_S2}. $e^-$ diffusion dominates over cation diffusion. Assuming this $D$ (and the $T=173$~K from earlier), as well as $\epsilon_d$ as defined above, and taking $v_d=1.51$~mm/$\mu$s at field $\mathcal{E}=180$~V/cm~\cite{luxrun3re2016}, we find $\alpha_{TI}=1.20\times10^{-9}$~m$^3$/s. From this, the escape probability ($1-\langle r \rangle$) for electrons inside a box is found by solving the relevant (Jaff\'e) differential equations (please see Section 6.2 of \cite{Dahl:2009nta} for the details).

We interpret $a_{TI}$, the size of the ``box'' surrounding ionized atoms, as corresponding to a ($\mathcal{E}$-independent) $e^-$-ion thermalization distance of 4.6~$\mu$m, as calculated by Mozumder~\cite{Mozumder_1995}. This value was used before as a border in NEST for track length, to switch from T-I to Birks. The ultimate value of TIB $\equiv\alpha_{TI}/(a_{TI}^2 v_d)$ for that case is 0.0376.

Dahl found best-fit values of TIB ranging from 0.03--0.04 for both ER and NR data at 60--522~V/cm~\cite{Dahl:2009nta}. Our contemporary fits (for NEST and for data), the blue lines at low energies in the first two panels at top in Figure~\ref{Fig4}, used 0.0300. If $v_d$ changes with drift field (it is typically $O$(2~mm/$\mu$s)~\cite{Albert:2016bhh}), then the entire ranges of Dahl, and of Sorensen and Dahl, are covered: 0.02--0.05~\cite{Sorensen_2011}.

\begin{figure}
\centering
\includegraphics[width=0.91\textwidth,clip, angle=0]{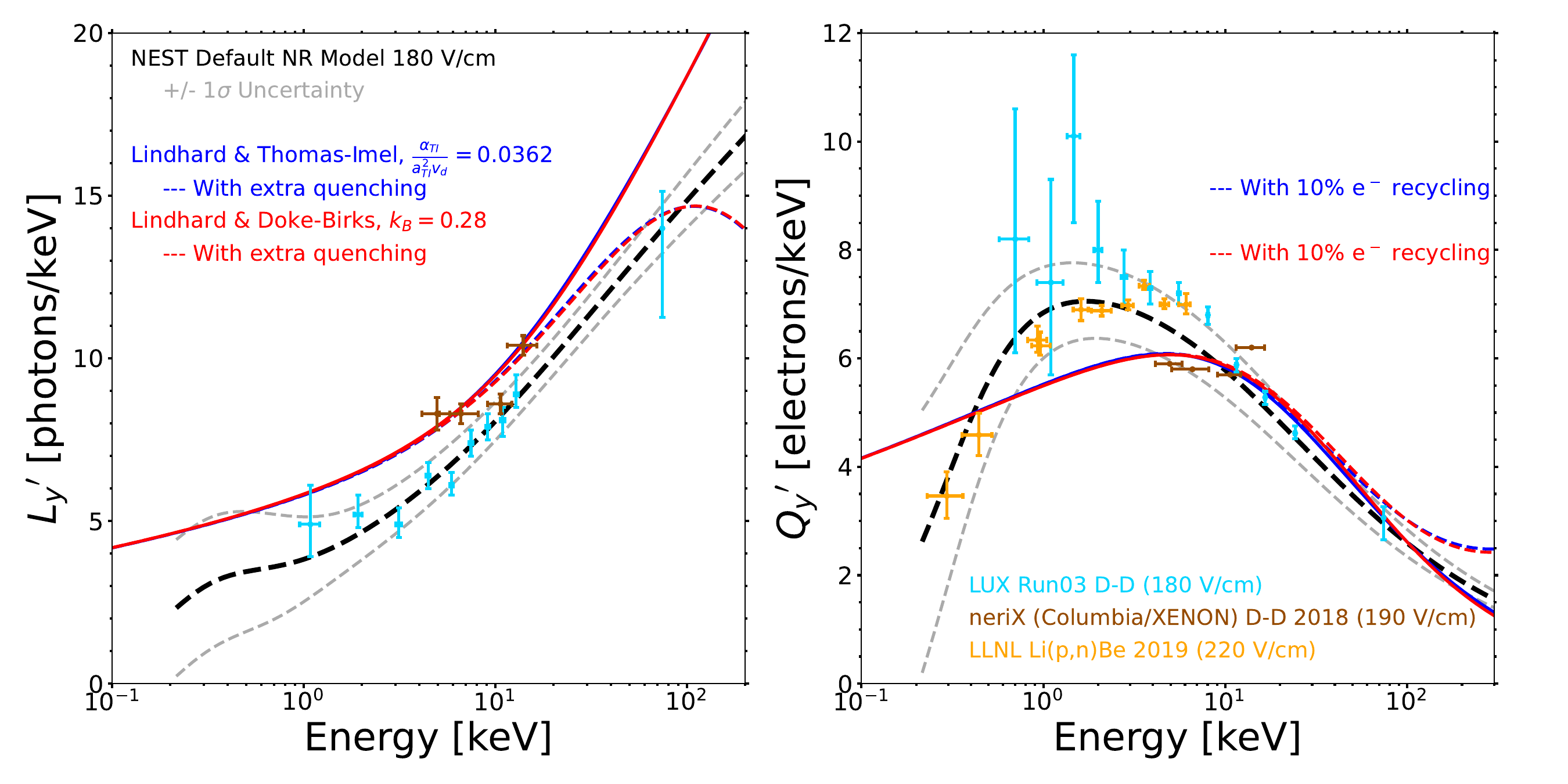}
\vspace{-15pt}
\caption{The comparisons of NEST and selected NR data to only the Thomas-Imel box (blue) and Birks (red) models of recombination, always using Lindhard here to define $N_q'$ (found as Eqn.~8 in \cite{instruments5010013} and elsewhere). For $L_y'$, the dashed lines indicate additional quenching at higher $E$s and $dE/dx$, while for the $Q_y'$, where this quenching has no direct impact, the dotted lines indicate partial conversion of photons into $e^-$s from that effect (or not, solid lines). Some data, including at other fields, are consistent at a 1--2$\sigma$ level with no quenching or conversion, not the amounts shown. The $L_y'$ data from 50--100~keV$_{nr}$ are inconsistent: see Figure~\ref{Fig3} upper left and \cite{PlanteLeff2011}.}
\label{Fig5}
\end{figure}

For NR, one sees in Figure~\ref{Fig4} (bottom row) many different past models, mainly for $L_y$. NEST originally used T-I for NR, as Dahl~/~Sorensen~\cite{Dahl:2009nta,Sorensen_2011}. See the blue lines in Figure~\ref{Fig5}. It applies the same color convention as Figure~\ref{Fig4}. While T-I fixes $r$, thus partitioning $E$ into $L_y$ vs.~$Q_y$, the total yield must still be determined. For the maximal distinction, we have selected the original Lindhard formula for that, as laid out in multiple other works~\cite{Lindhard_1963,Sorensen_2011,LUXDD2015,instruments5010013}, not Equation~\ref{eqn:9}. We set the crucial Lindhard parameter $k_L$ = 0.166, the decades-old default for Xe~\cite{Lindhard_1963}. Averaging over $E$, $N_q'/N_q \approx k_L$. 0.166 is consistent with actual data~\cite{LUXDD2015}, Lenardo's meta-analysis~\cite{NEST2015}, and NEST v2.3+.

We identify $\varsigma$ of Equation~\ref{eqn:12} with TIB value, as justified by Equation~\ref{eqn:11}, wherein the parameters for the $\mathcal{E}$ dependence of $\varsigma$ ($\gamma$ and $\delta$) overlap at the 1$\sigma$ level with the power-law field dependence of TIB from \cite{NEST2015}. At 180 V/cm, $\varsigma=0.0362$, quite close to our theoretical calculation earlier and comparable to a best-fit TIB for ER. $N_{ex}/N_i=1.0$ is assumed. While higher than for ER, it is the most common assumption for NR, and best fits to data and theory vary from 0.7--1.1~\cite{Sorensen_2011}.

An additional quenching is applied to just $L_y'$~\cite{Manzur:2009hp}. We find a common parameterization of this effect~\cite{Bezrukov_2011} to be defined in a manner analagous to Birks' Law or Equation~\ref{eqn:14}:

\begin{equation}
q = \frac{1}{1 + \kappa \epsilon_Z ^{~~\lambda}},~\mathrm{with}~\epsilon_Z \approx 10^{-3} E,
\label{eqn:16}
\end{equation}

\noindent
where $q < 1$ is a multiplicative factor on $L_y'$. $\epsilon_Z$ is unitless reduced energy, useful for comparison between elements. Equation~\ref{eqn:16} is like \ref{eqn:14}. The power law can be identified as proportional to NR $dE/dx$. If we define $dE/dx$ (or LET) as approximately $\beta' \epsilon ^ \lambda$, then $\kappa = k_B \beta' \mathcal{L}$. Assuming the ER $k_B$ (defined as 0.28 for 180~V/cm in Figure~\ref{Fig4} top), $\mathcal{L} \sim 0.15$ (11/73) per an energy-independent approximation of Equation~\ref{eqn:9} justified by the power being close to 1, and $\beta'=100$, then $\kappa = 4.20$, $<0.2\sigma$ away from \cite{NEST2015}. A fraction of the quanta removed from $L_y'$ in \ref{eqn:16} may be convertible into $Q_y'$. Figure~\ref{Fig5} right explores that with the fraction as 0.1.

Unlike with ER, Birks' Law models NR over the entire $E$ range of interest (Figure~\ref{Fig5}, red) with $k_B = 0.28$ and $dE/dx = \beta' \epsilon ^ \lambda = 100 \epsilon$. While there is disagreement about whether $\lambda$ is 1.0 or 0.5 depending on the $E$ regime \cite{Hitachi,PhysRevLett.97.081302}, 1.0 only differs by $1.6\sigma$ from the value of 1.14 in \cite{NEST2015}.

Looking back at alternatives to Lindhard, in Figure~\ref{Fig4}, we see NEST's power law, $L_y'$, and $Q_y'$ seem a good match for Mu and Xiong~\cite{mu2013scintillation,mu2013ionization}, also for Wang and Mei~\cite{Wang:2016obw,Mei_2008}. NEST's lower $1\sigma$ line touches Sarkis' $L_y'$~\cite{Sarkis_2020}, which is low due to not including the most recent points~\cite{LUXDD2015,DQLUXRun04DD}. On the high-$E$ $L_y'$ end, NEST's upper uncertainty band encompasses neriX~\cite{Aprile_2018_NeriXNR}. As for $Q_y'$, NEST lies in between \cite{Wang:2016obw} above and \cite{mu2013ionization} and \cite{Sarkis_2020} below, falling in between LUX D--D~\cite{LUXDD2015} and LLNL~\cite{lenardo2019measurement}.

The good agreement between the fully empirical NEST model and the first-principle models of both NR and ER shown here shows that NEST can accurately simulate potential dark matter signals and backgrounds, respectively. This should be the case even for the regimes where data are still lacking, or they exist but have large uncertainties. In the case of NR, the fully empirical approach reproduces all data better using a comparable number of free parameters, but much greater flexibility compared to semi-empirical approaches. For fluctuations, the number of NEST free parameters increased to two Fano factors (excitation, ionization) and four numbers for recombination width and skew, to fully model the $E$ resolution. NEST, justified first using data, is not limited to the operating conditions of previous experiments to make predictions relevant for a future experiment, though $L_y$ and $Q_y$ must pass through a detector simulation to obtain realistic S1 and S2 pulse areas: the processes in Section 2.2.5 here and Appendix A of \cite{James_2022}.

\vspace{-25pt}
\section{Discussion and Future Work}
\vspace{-5pt}
Beginning with our models of beta ER, gamma-ray ER, and the NR light and charge yields, along with resolution modeling, a coherent picture was built up inside of the NEST framework, which enables a good agreement with data. NEST was also shown to have features from multiple first-principles approaches, such as the box and Birks models. NEST already works for LAr~\cite{instruments5010013} using the same formulae as LXe but unique parameter values. However, it still works best only for point-like interactions, like those in dark matter experiments like DarkSide, not tracks, as will be observed by DUNE. The list of NEST collaborators includes TESSERACT~\cite{Biekert_2022} members, so addition of liquid helium (LHe) to NEST is planned.

Looking beyond LHe, short-term future work includes a NEST re-writing to account for the lower $W_q$ measured by EXO and Baudis et al.~\cite{Anton:2019hnw,Baudis_2021_NewW}, but this will be easier if NEST can return to approaches closer to first principles. Therefore, there will also be a concerted effort made to return to a semi-empirical formulation by application of a modified T-I model pioneered by ArgoNeuT~\cite{Acciarri_2013}, combined with a literal breakup of long tracks into boxes as done in the thesis of Dahl, allowing higher energies to exhibit lower light yields without hard-coding that, by virtue of being comprised of multiple, lower-$E$ interaction sites. High-$E$ modeling is thus accomplished by having one model for all $E$s, but treating high-$E$ interactions as series of many low-$E$ fragments wherein the $L_y$ will continue to be monotonically increasing with $E$. The main motivation for this is greater confidence in extrapolations to uncalibrated regions of future detectors.

The modified box model of LArTPC-based high-$E$ neutrino experiments should also be useful for LXe NR. We demonstrate now how it represents a more generalized version of the current NR model:

\begin{equation}
Q_y'=\frac{N_{e-}}{E}=(1-r)\frac{N_i}{E}=\frac{\mathrm{ln}(a'+\xi)}{\xi}\frac{N_i}{E}=\frac{\mathrm{ln}(a'+\xi)}{\xi}\frac{N_q/E}{1+\alpha_x}=\frac{\mathrm{ln}(a'+\xi)}{\xi E}\frac{aE^b}{1+\alpha_x},
\label{eqn:17}
\end{equation}

\noindent
where $a' \equiv 1$ in default T-I (but relaxing this constraint to $O$(1) as per \cite{Acciarri_2013} can better fit data), $\xi$ is short for $\xi_{TI}$, redefined as $\beta~dE/dx$ with $\beta$ as a constant (not Equation~\ref{eqn:15}), and $\alpha_x \equiv N_{ex}/N_i$ for conciseness.

\begin{equation}
Q_y'=\frac{aE^{b-1}}{1+\alpha_x}\frac{\mathrm{ln}(1+\beta\frac{dE}{dx})}{\beta\frac{dE}{dx}} \approx \frac{a}{1+\alpha_x}\frac{\mathrm{ln}(1+\beta\frac{dE}{dx})}{\beta\frac{dE}{dx}} \approx \frac{a}{2}\frac{\mathrm{ln}(1+\beta\frac{dE}{dx})}{\beta\frac{dE}{dx}} \approx 5 \frac{\mathrm{ln}(1+\beta\frac{dE}{dx})}{\beta\frac{dE}{dx}},
\end{equation}

\noindent
where we employ, in order, the approximations $b \approx 1$, $\alpha_x \approx 1$, and $a \approx 10$ (Equation~\ref{eqn:9}). Fitting to the SRIM line in Fig.~5 of \cite{PhysRevLett.97.081302} one finds for NR, in normalized (dimensionless) units, that stopping power is:

\begin{equation}
dE/dx = \beta' \epsilon_Z ^{0.5} = 120\sqrt{\epsilon_Z} = 120\sqrt{0.001E} = 120\sqrt{0.001}\sqrt{E} = 3.8\sqrt{E},
\end{equation}

\noindent valid for the range of 0--100~keV. However, near 50~keV, a square root function with an offset fits nearly as well to SRIM: $3.4\sqrt{E+\epsilon}$, with $\epsilon$ = 12.6~keV (Equation~\ref{eqn:12}). Making the ansatz $\beta \sim \varsigma$ (Equation~\ref{eqn:11}),

\begin{equation}
Q_y' \approx 5 \frac{\mathrm{ln}(1+\varsigma\frac{dE}{dx})}{\varsigma\frac{dE}{dx}} = 5 \frac{\mathrm{ln}(1+0.036~3.4~\sqrt{E+12.6})}{0.036~3.4~\sqrt{E+12.6}} = \frac{5}{3.4} \frac{0.677}{\varsigma\sqrt{E+\epsilon}} = \frac{1}{\varsigma\sqrt{E+\epsilon}},
\label{eqn:20}
\end{equation}

\noindent
recovering the high-$E$ portion of Equation~\ref{eqn:12} at $\varsigma=0.036$ (200~V/cm) and $E$ = 50~keV. By modifying the power law for $N_q$ to be $aE^b-C$~\cite{McMThesis}, it may be possible to eliminate the need for the sigmoids for reducing both $Q_y$ and $L_y$ at the lowest $E$s, combining $C$ with an additional degree of freedom, a non-unity $a'$ in the natural log. By replacing our present equation (\ref{eqn:12} or \ref{eqn:20}) with Equation~\ref{eqn:17} we should be able to find a sufficiently-flexible compromise that fits data with the same number of free parameters, or fewer even (eliminating the sigmoid roll-offs, as well as the $\epsilon$ offset in $dE/dx$ potentially) all motivated from first principles (T-I). The redefinition of $\xi_{TI}$ in terms of $dE/dx$ permits a non-linearity in the dependence of $\xi$ on $N_i$ and an incorporation of $dE/dx$ (as in the Doke~/~Birks' Law), while $\alpha_x$ could be made $E$ and $\mathcal{E}$-dependent as in Eq.~(B8) of \cite{xenoncollaboration2024xenonntwimpsearchsignal}, if absolutely necessary, following the similar increase with $E$ for ER in Equation~\ref{eqn:2} (mimicked by Eq.~(A4)'s exponential in \cite{xenoncollaboration2024xenonntwimpsearchsignal}). Lastly, the replacement of $aE^b-C$ with $E/W_{q}$ in Equation~\ref{eqn:17} could permit usage for ER, as in LAr, from the keV to the GeV scales.

Improved modeling of the MeV (ERs) scale is important for searches for neutrinoless double-beta ($0\nu\beta\beta$) decay, for which the key discrimination is not NR vs.~ER, but between two forms of the latter ($\beta$ vs.~$\gamma$). EXO-200~\cite{Anton:2019wmi} and KamLAND-Zen~\cite{KZ} have produced the two most stringent half-life limits for $^{136}$Xe, and are highly competitive with the Ge-based experiments. In addition to these results, one must evaluate the prospects of nEXO~\cite{nEXO}, as well as of LZ~\cite{Akerib_2020_0vBB}, XENONnT~\cite{PhysRevC.106.024328}, and XLZD~\cite{NextGen} for this field of nuclear physics. The dark-matter-focused experiments have greater ER backgrounds compared to nEXO, but superior energy resolution.

Longer-term future work on NEST will involve an \textit{ab initio} MC with cross sections for recombination and the other processes~\cite{OliviaDanVetri}, and/or molecular dynamics modeling of Xe atoms with the 12-6 Lennard-Jones potential for van der Waals forces. LXe values are known, for L-J (and more advanced models~\cite{Gabor}):
\vspace{-1pt}
\begin{equation}
V(d)= 4 \epsilon_{LJ} \left [ {\left (\dfrac{R}{d} \right )}^{12}-{\left (\dfrac{R}{d} \right )}^{6} \right],~\mathrm{where}~\epsilon_{LJ}=1.77~\mathrm{kJ/mol}~\mathrm{and}~R=4.10~\textup{~\AA}
\label{eqn:21}
\end{equation}

\noindent
While these approaches are challenging for high (MeV) energies, at sub-keV scales where yields are more uncertain, \textit{e.g.}~for $^8$B, fewer interactions are involved, leading to a more computationally tractable problem.

\section*{Conflict of Interest Statement}

All of the authors declare that the research was conducted in the absence of any commercial or financial relationships that could be construed as a potential conflict of interest.

\section*{Author Contributions}

MS and GRCR did most of the writing and are responsible for the majority of plots and analyses. JB forged the original beta models; GAB worked on ER energy reconstruction; JPB debugged code; EB, ACK, and DNM read and revised text; JEC and ESK improved the functionality of the NEST code and contributed to multiple unique analyses; SJF and CSL created Figures 1--3 together; JH and ZZ worked on the gamma model; KM researched the work function discrepancy; RM, KT, and MDW worked on the NR models; MM, JM, and SW provided an outside perspective, from the LAr communities; KN and CDT represented the XENON collaboration's points of view; MT hosted the NEST website at his institution (UCD); VV read the text and revised it extensively, in addition to being responsible for Figures 4 and 5. MZ looked at the field dependence of the ER and NR yields.

\section*{Funding}

This work was supported by the U.S.~Department of Energy (DOE) under Awards DE-SC0015535, DE-SC0024225, DE-SC0021388, DE-SC0018982 and DE-AC02-05CH11231, and by the National Science Foundation (NSF) under Awards 2046549 and 2112802.

\section*{Acknowledgments}

We thank the LZ/LUX plus XENON1T/nT/DARWIN collaborations for useful recent discussion as well as continued support for NEST work. We especially thank LUX for providing key detector parameters, and LUX collaborator Prof.~Rick Gaitskell (of Brown University), Dr.~Xin Xiang (formerly of Brown, now at Brookhaven National Laboratory), and Dr.~Quentin Riffard (Lawrence Berkeley National Laboratory), for critical discussions regarding the detector performance of a potential Generation-3 liquid Xe TPC detector.

\section*{Supplemental Data}
Please see Appendices A and B.
\section*{Data Availability Statement}
The data presented in this study are available upon request.

\bibliographystyle{Frontiers-Vancouver}
\bibliography{nest}
\clearpage

%\documentclass[utf8]{frontiers_suppmat}
%\usepackage{url,hyperref,lineno,microtype,silence}
%\usepackage[onehalfspacing]{setspace}
%\WarningFilter{microtype}{Unable to apply patch `footnote'}
%\linenumbers

%\begin{document}
%\onecolumn
%\firstpage{1}

%\title[Supplementary Material]{{\helveticaitalic{Supplementary Material}}}
\section*{Supplementary Material}

%\maketitle
\subsection*{Appendix A: Work Function, Exciton-Ion Ratio, and Fano Factor}

\begin{figure}[ht!]
\includegraphics[width=\textwidth]{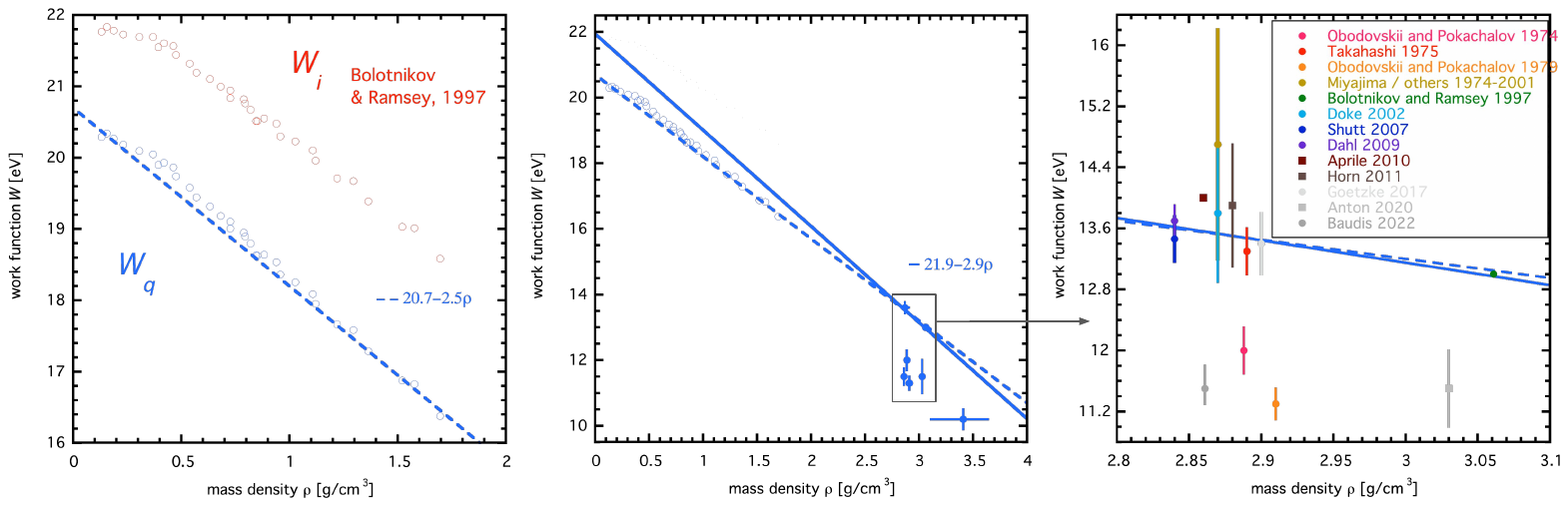}
\vspace{-191pt}
\caption{Left: The ionization work function, $W_i$, as a function of the density for GXe (gaseous Xe)~as hollow red circles~[\citenum{Bolotnikov}] normalized to $W_{i,0} = 21.8$~eV, and converted to $W_q$ in cyan based upon $N_{ex}/N_i = 0.0674 + 0.0397 \rho$, the simplest way to account for a value of 0.06--0.07 for room-temperature/pressure gas~[\citenum{Miyajima:1974zz}] and the highest fit value for LXe at the highest $E$s in LXe, 0.20~[\citenum{Doke_2002}]. That linear dependence of $N_{ex}/N_i$ on $\rho$ also serves to flatten the curvature of the $W$ dependence, allowing for a linear fit, though with densities between 0.5 and 1 still ``bulging'' past the line (NEST previously utilized a sigmoid but abandoned it for simplicity, given the unknown errors on the data and an \textit{ad hoc} exciton-ion ratio dependency.) Middle: the same corrected data repeated in cyan and same fit, but with data from condensed Xe (liquid and solid) added as solid circles. The only known data point for solid Xe~[\citenum{OboPok1974}] is the lowest / right-most, at an unknown $\rho$ (taken to be 3.41~[\citenum{Yoo:2015xza}], but with an uncertainty spanning 3.1--3.64). The high-$W_q$ point above 3~g/cm$^3$ without an error is part of the same [\citenum{Bolotnikov}] data set as the GXe points. A new two-point fit to only LXe data points, those agreeing on a higher $W_q$ near 2.88 averaged together, and the single point at 3.06, is introduced as a solid cyan line (the original repeated as a dashed line). Due to the uncertainty in the normalization of the hollow circles (GXe) the new steeper line may still agree with them. Right: A blow-up of LXe data points, with all high-$W_q$ points below 3~g/cm$^3$ that were merged (error-weighted average) into one point in the middle plot now separated, broken down by source in the legend: pink [\citenum{OboPok1974}], red [\citenum{Takahashi1975}], orange [\citenum{OboPok1979}], yellow [\citenum{Miyajima:1974zz,DOKE1990617,DOKE1993113,TANAKA2001454}]. The remaining, from top to bottom, are: [\citenum{Bolotnikov,Doke_2002,SHUTT2007451,Dahl:2009nta,bib:37,Horn_2011,Goetzke:2016lfg,Anton:2019hnw,Baudis_2021_NewW}].}
\label{fig:1}
\vspace{-5pt}
\end{figure}

This appendix presents the origins of the density-dependent work function for total quanta $W_q$ and the high-$E$ asymptote of the exciton-to-ion ratio $N_{ex}/N_i$ implemented within NEST, as well as the Fano-like factor for total quanta, all for ER. Note that Figure~\ref{fig:1} uses a 21.8~eV normalization for $W_i$, reported as going to 1.00 as $\rho \to 0$~g/cm$^3$ in [\citenum{Aprile:2008bga}] and [\citenum{Bolotnikov}]. That value is based upon averaging three values reported in the same sources without recorded error bars: 21.5 and 22.1 (originally from [\citenum{Ahlen}], and also the ICRU value) in the former source and 21.9 in the latter. It represents 1~/~$Q_y$ as $\mathcal{E}$ $\to$ $\infty$. ($W_q$ $<$ $W_i$ by definition.)

\vspace{-71pt}

A reduction in $W_q$ with density and between the gaseous and condensed phases is observed in nearly all noble elements~[\citenum{Aprile:2008bga}], meaning it takes less $E$ to either excite or ionize atoms as they get closer together. Additionally, $N_{ex}/N_i$ appears to increase with $\rho$. This suggests that as mass density increases ionization density does as well, and recombination becomes stronger, resulting in a component of it that is near-immediate and can be modeled as direct excitation. This is known as Onsager or geminate recombination, and is effectively the opposite of volume or columnar recombination~[\citenum{Nygren_2013}]. Another possibility, as mentioned in Section~2.2, is that $W_i$ and $W_{ex}$ must be considered separately. That would complicate the definition of a combined-$E$ scale for energy reconstruction, useful for any phase. It relies on strict anti-correlation.

\begin{figure}[ht!]
\includegraphics[width=\textwidth]{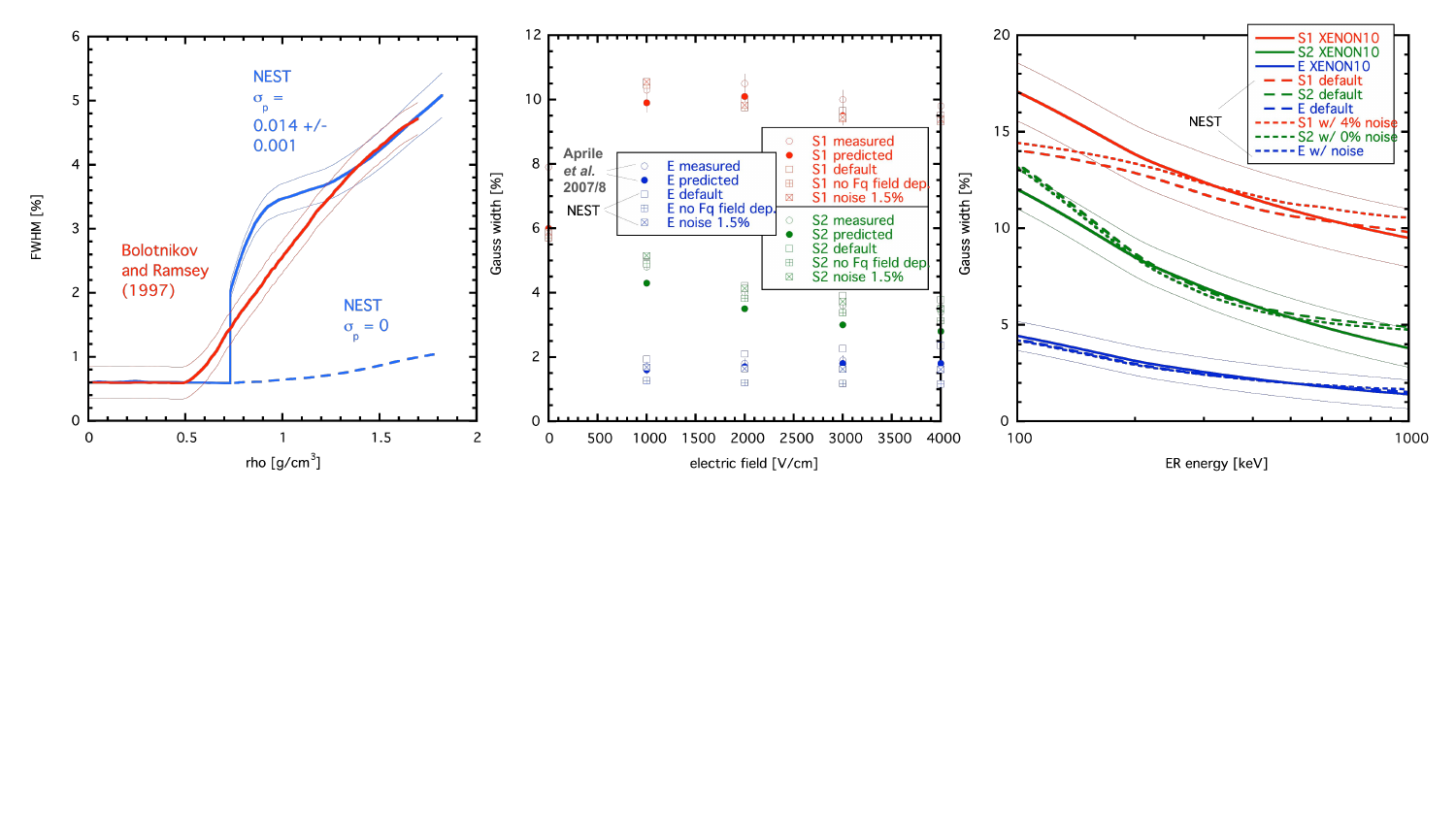}
\vspace{-140pt}
\caption{Left: The ionization-only resolution of $^{137}$Cs gamma rays in GXe at 7~kV/cm vs.~density~[\citenum{Bolotnikov}]~in red, with approximate error band (note: these same data feature prominently in [\citenum{NYGREN2009337}]). At very low densities there is nearly no recombination: S1 comes only from excitations at this very high field, but $N_{ex}/N_i$ is low too (near 0.06). Thus, resolution in this flat region is driven almost exclusively by low $F_q = 0.1-0.2 \lessapprox F_i$. To follow $F$'s $\rho$ dependence, featured in Eq.~(7) in the main text, a cubic spline was fit to discrete $F$s measured or predicted (calculated) for xenon as a low-pressure gas, supercritical fluid, liquid, and solid, featured in Table 2.4 of [\citenum{Aprile:2008bga}]. NEST (cyan) was intentionally not fit to the red, to see if it could be predicted. As a result it does not quite agree on the $\rho$ at which resolution begins to depart from flatness, at $O(100)$ bar, and has some disagreement at moderate $\rho$s (though no actual major experiments operate under such conditions any longer). The dashed cyan line represents only binomial recombination fluctuations, with the closest fit to real data achieved with a non-zero value of the non-binomial contribution (solid lines) thus demonstrating that these matter even in a gas or a van der Waals fluid not just true liquid, at sufficiently high densities. Energy resolution is not driven by just $F_q$, when considering only a single channel such~as ionization alone. Middle: Resolution as a function of $\mathcal{E}$ for LXe (2.9~g/cm$^3$) for the same fixed $E$ = 662~keV and particle ($\gamma$) again, for both ionization (S2, green) and scintillation (S1, red) now, as well as for the combined-$E$ scale (blue) that merges the information from both to create the best-possible resolution / lowest width~[\citenum{Aprile_2007}]. Default NEST options reproduce real data well, but without the $F_q$ square-root $\mathcal{E}$ dependence from Eqn.~(7) in the text the final resolution is seriously underestimated: 1.1--1.2 compared to 1.5--2.0\% in reality. Another option for matching data with NEST is presented where $F_q \approx 0$ and linear noise terms are added to represent unknown detector effects, which the ``predicted'' points from Aprile were meant to address. In all NEST options presented, $g_1 = 0.05$, typical for the time of data collection, and 40~V/cm stands in for 0~V/cm (impossible to directly model within NEST, as explained in the text). Note that [\citenum{Bolotnikov}] also has LXe data, and found 2.9\% (converted from FWHM) resolution at 7000~V/cm, in good agreement with the asymptotic behavior of the S2s seen in this plot. Right: With density (LXe) and field (730~V/cm) both fixed now this is the $E$ dependence, for XENON10~[\citenum{bib:37}]. Two NEST options are presented: the default, plus usage of Gaussian noise terms again, at $F_q = 0.03$ (no $\mathcal{E}$ nor $E$ dependence, only $\rho$). At 662 keV, the measured S1, S2, and total ($E$) resolutions were 10.8, 5.1, and 2.2\%, respectively, all comparable to the 1~kV/cm results from the middle plot.}
\label{fig:2}
\vspace{-20pt}
\end{figure}

Next, we cover the Fano-like factor $F_q$ for variation in total quanta in greater depth than in the main body of the text, plus the excitation / recombination fluctuations again. For the former, the origins of its density, field, and $E$ dependences are presented in Figure~\ref{fig:2} from left to right. Resolution here is defined as Gaussian width divided by median. The full width half max (FWHM) is used for the first plot, as that is how its data were originally reported, and standard deviation is used for the others. Results from $^{137}$Cs are displayed, one of the most common past standard candles after $^{57}$Co. Xe gas and supercritical fluid at room temperature but distinct pressures are represented in the left pane (up to $>$ 60 bar, corresponding to about 1.8~g/cm$^3$) while the other two are for liquid, where the standard NEST recombination fluctuations for LXe ($\sigma_p$ = 0.04--0.09) are applied, dependent on $\mathcal{E}$ and $E$.

The recombination fluctuations, which become canceled out on a combined-$E$ scale, were historically often conflated with $F_q$. Effective $F_{ph}$ and $F_{e-}$ (aka $F_{sc}$ and $F_{i}$) can still be defined, with enormous but inconsequential values: 60 and 20 for the $^{137}$Cs example, at 1000 V/cm in LXe, matching [\citenum{NYGREN2009337}].

\vspace{-15pt}

\subsection*{Appendix B: Tabulation of NEST Model Parameters}
\vspace{-5pt}
In this appendix, we provide tables detailing the functions and model parameters used in NEST for LXe yields from $\beta$ ER, $\gamma$ ER, NR, as well as their fluctuations. NEST has additional models for ${}^{83m}$Kr ER as well as NR from non-Xe nuclei (including $\alpha$ decay), findable in code on GitHub [\citenum{szydagis_m_2023_8215927}].

\vspace{-20pt}

\begin{table*}[h!]
\caption{\footnotesize{Table of NEST model parameters comprising the $\beta$ ER yield models for charge, as shown in Equation~(5), and light.}}
\vspace{-0pt}
 \centering
    \begin{tabular}{ | P{0.15\linewidth} | p{0.8\linewidth}|}
\hline
    $m_1$ & Stitching-region yield for $\beta$ ER charge yields between low and high energies, depending on field and density: $m_1 = 30.66 + (6.20 - 30.66) / (1 + (\mathcal{E} / 73.86)^{2.03})^{0.42}$ at a typical LXe density. Takes values $\mathcal{O}$(10 keV$^{-1}$) for $\mathcal{O}$(100~V/cm) fields. \\ \hline
    $m_2$ & Low-energy asymptote of the $\beta$ ER charge yield equation. Default value is approximately 77.3~keV$^{-1}$. \\   \hline
    $m_3$ & Controls the energy-dependent shape of the $\beta$ charge yields in the low-energy (Thomas-Imel) regime: $m_3 = \text{log}_{10}(\mathcal{E}) \cdot 0.14 + 0.53$. Field-dependent function, with values of approximately 0.8-1.5~keV for $\mathcal{O}$(100~V/cm) fields. \\ \hline
    $m_4$ & Field-dependent control on the energy-dependent shape of the $\beta$ charge yields at lower energies: $m_4 = 1.82 +  (2.83 - 1.82) / (1 + (\mathcal{E} / 144.65)^{-2.81})$. Takes values from approximately 2.0-2.8 for $\mathcal{O}$(100~V/cm) fields. \\ \hline
    $m_5$ & High-energy asymptote of the $\beta$ charge yield model. Defined as: $m_5$ = $\frac{1}{W}\cdot[1 + N_{ex}/N_i]^{-1} - m_1$ (See Ref.~[\citenum{GregRC14}].) \\ \hline
    $m_6$ & Low-energy asymptote of the higher-energy behavior for $\beta$ ER charge yields. Degenerate with $m_1$ and explicitly set to 0~keV$^{-1}$. \\ \hline
    $m_7$ & Field-dependent scaling on the behavior of the $\beta$ charge yields at higher energies: $m_7 = 7.03 + (98.28 - 7.03) / (1. + (\mathcal{E} / 256.48)^{1.29})$. Takes values $\mathcal{O}$(10 keV) for $\mathcal{O}$(100~V/cm) fields. \\ \hline
    $m_8$ & Control on the energy-dependent shape of the $\beta$ charge yields at higher energies. The default value is a constant, 4.3. \\ \hline
    $m_9$ & Asymmetry control on the low-energy behavior. The default value is a constant, 0.3. \\ \hline
    $m_{10}$ & Asymmetry control on the high-energy behavior of the $\beta$ charge yields model: $m_{10} = 0.05 + (0.12 - 0.05)/(1 + (\mathcal{E} / 139.26)^{-0.66})$. Field-dependent function that takes values $\sim$0.1 for $\mathcal{O}$(100~V/cm) fields. \\ \hline
    \end{tabular}
\end{table*}

\vspace{-20pt}

\begin{table*}[hb!]
\caption{\footnotesize{Table of NEST model parameters comprising the $\gamma$ ER yield models for light and for charge, reusing Equation~(5) from the $\beta$ ER yields.}}
\vspace{-0pt}
 \centering
    \begin{tabular}{ | P{0.15\linewidth} | p{0.8\linewidth}|}
\hline
    $m_1$ & Field-dependent function controlling the transition between lower and higher energies: $m_1 = 34.0 + (3.3 - 34.0)/(1 + (\mathcal{E}/165.3)^{0.7}$). \\ \hline
    $m_2$ & Low-energy asymptote of the $\gamma$ ER charge yield equation, defined as 1/$W_q$ in units of keV$^{-1}$. \\   \hline
    $m_3$ & Controls the energy-dependent shape of the $\gamma$ charge yields in the low-energy (Thomas-Imel) regime; a constant value of 2~keV is used. \\ \hline
    $m_4$ & Control on the energy-dependent shape of the $\gamma$ charge yields at lower energies; a constant power of 2 is used. \\ \hline
    $m_5$ & High-energy asymptote of the $\gamma$ charge yield model. Defined as: $m_5 = 23.2 + (10.7 -23.1)/(1 + (\mathcal{E}/34.2)^{0.9})$. \\ \hline
    $m_6$ & Low-energy asymptote of the higher-energy behavior for $\gamma$ ER charge yields. Degenerate with $m_1$ and explicitly set to 0~keV$^{-1}$. \\ \hline
    $m_7$ & Field-dependent and density-dependent scaling on the behavior of the $\gamma$ charge yields at higher energies: $m_7 = 66.8 + (829.3 - 66.8)/(1 + (\rho^{8.2}\cdot\mathcal{E}/(2.4\cdot10^5))^{0.8})$.  \\ \hline
    $m_8$ & Control on the energy-dependent shape of the $\gamma$ charge yields at higher energies. Default value is a constant power of 2. \\ \hline
    $m_9$ & Asymmetry control on the low-energy behavior: unused for $\gamma$ ER yields and set to unity. \\ \hline
    $m_{10}$ & Asymmetry control on the high-energy behavior of the $\gamma$ charge yields model: unused for $\gamma$ ER yields and set to unity. \\ \hline
    
    \end{tabular}
\end{table*}

\vspace{-5pt}

\begin{table*}[ht!]
\caption{Table of NEST model parameters comprising the NR mean yield models: for total quanta, charge, and light, as shown in Equations (9), (12), and (13).}
\vspace{-0pt}
 \centering
    \begin{tabular}{ | P{0.15\linewidth} | p{0.8\linewidth}|}
\hline
    $a$ & Scaling on NR total quanta. Default value is 11${}^{+2.0}_{-0.5}$~keV$^{-b}$. \\ \hline
    $b$ & Power-law exponent for the NR total quanta. Default value is 1.1 $\pm$ 0.05. \\ \hline
    $\varsigma$ & Field dependence in NR light and charge yields, with mass-density-dependent scaling (Equation (11)). \\ \hline
    $\rho_0$ & Reference density for scaling density-dependent NEST functions: 2.90 g/cm$^3$. \\ \hline
    $v$ & Hypothetical exponential control on density dependence in $\varsigma$; the default value is 0.3. \\ \hline
    $\gamma$ & Power-law base for the field dependence in $\varsigma$. Default value is 0.0480 $\pm$ 0.0021. \\ \hline
    $\delta$ & Power-law exponent in the field dependence in $\varsigma$; default value is -0.0533 $\pm$ 0.0068. \\ \hline
    $\epsilon$ & Reshaping parameter for NR charge yields, controlling the effective energy scale at which the charge yield behavior changes. The default value is 12.6${}^{+3.4}_{-2.9}$~keV. \\ \hline
    $p$ & Exponent which controls the shape of the energy dependence of the NR charge yields at energies greater than $\mathcal{O}(\epsilon)$. Default value is 0.5. \\ \hline
    $\zeta$ & Controls the energy dependence of the NR charge yields roll-off at low energies. Default value is 0.3 $\pm$ 0.1 keV.\\ \hline
    $\eta$ & Controls energy-dependent shape of the NR charge yields roll-off at low energies. Default value is 2 $\pm$ 1. \\ \hline
    $\theta$ & Controls the energy dependence of the NR light yields roll-off. Default value is 0.30 $\pm$ 0.05 keV. \\ \hline
    $\iota$ & Controls the shape of the energy dependence of the NR light yields roll-off. Default value is 2.0 $\pm$ 0.5. 
    \\ \hline
\end{tabular}
\vspace{-6pt}
\end{table*}

\begin{table*}[ht!]
\caption{Table of NEST model parameters for different types of fluctuations for ERs and NRs.}
\vspace{-0pt}
 \centering
    \begin{tabular}{ | P{0.15\linewidth} | p{0.8\linewidth}|}
\hline
    $F_q$ & Fano-like factor for statistical fluctuations. For ERs, this is proportional to $\sqrt{E \cdot \mathcal{E}}$; see Equation~(7). For NRs, this is separated into fluctuations for $N_{ex}$ and $N_{i}$; the default value is 0.4 for both in NEST v2.3.11, while the values were 1.0 in previous NEST versions. ($F_{ex}$ was underestimated to be conservative for low-mass WIMPs.) \\ \hline
    $\sigma_p$ & Non-binomial contribution to recombination fluctuations, modeled as a skew Gaussian in electron fraction space. \\ \hline
    $A$ & Amplitude of non-binomial recombination skew Gaussian. For NRs, this is a constant 0.04 (v2.3.11) or 0.1 (v2.3.10). For ERs, it is field-dependent: $A = 0.09 + (0.05 - 0.09)/(1 + (\mathcal{E}/295.2)^{251.6})^{0.007})$, where 0.05 was 0.055 in 2.3.10 \\ \hline
    $\xi$ & Centroid-location parameter of the non-binomial recombination skew Gaussian. Default value for ERs is an electron fraction of 0.45, but 0.5 for NRs. \\ \hline
    $\omega$ & Width parameter for the non-binomial recombination skew Gaussian. Takes value of 0.205 for ERs and 0.19 for NRs. \\ \hline
    $\alpha_p$ & Skewness parameters for the non-binomial recombination skew Gaussian. Takes the value -0.2 for ERs, while being zero for NRs. \\ \hline
    $\alpha_r$ & Additional skewness in the recombination process itself. Field- and energy- dependent equations can be found in Ref.~[\citenum{VelanDiscrim}] for ERs, while this is fixed at 2.25 for NRs, with evidence of higher values in [\citenum{VelanDiscrim}]. \\ \hline
\end{tabular}
\end{table*}

%\bibliographystyle{Frontiers-Vancouver}
%\bibliography{nest}
%\end{document}
\end{document}